\documentclass[]{aa}
\usepackage{graphicx}
\usepackage{txfonts}
\begin{document}
   \title{The Parkes \ion{H}{i} Survey of the Magellanic System}

   \subtitle{}

   \author{C. Br\"uns\inst{1} 
          \and
   J. Kerp\inst{1} 
          \and
   L. Staveley-Smith\inst{2}
          \and
   U. Mebold\inst{1} 
   \and
   M.E. Putman\inst{3}
          \and
   R.F. Haynes\inst{2}
          \and
   P.M.W. Kalberla\inst{1}
          \and
          E.~Muller\inst{4}
	  \and
	  M.D. Filipovic\inst{2,5}
          }

   \offprints{C. Br\"uns, \\
   \email{cbruens@astro.uni-bonn.de}}

   \institute{Radioastronomisches Institut, Universit\"at Bonn,
   		Auf dem H\"ugel 71, D-53121 Bonn, Germany
         \and
             Australia Telescope National Facility, CSIRO,
       PO Box 76, Epping NSW 1710, Australia
	 \and
	 Department of Astronomy, University of Michigan, Ann Arbor, MI 48109, USA   
	 \and
	 Arecibo Observatory, HC3 Box 53995, Arecibo, PR 00612, USA
	 \and
	 University of Western Sydney, Locked Bag 1797, Penrith South, DC, NSW 1797,
	 Australia
             }

   \date{Received  24 February, 2004; accepted 27 October, 2004}

   \abstract{
We present the first fully and uniformly sampled, spatially complete \ion{H}{i} 
survey of the entire Magellanic System with high velocity resolution 
($\Delta v$ = 1.0~km\,s$^{-1}$), performed with the Parkes Telescope\thanks{The 
Parkes Telescope is part of the Australia Telescope which is funded by the 
Commonwealth of Australia for operation as a National Facility managed by CSIRO}. 
Approximately 24 percent of the southern sky was covered by this survey on a 
$\approx$ 5\arcmin\ grid with an angular resolution of HPBW = 14\farcm1. A 
fully automated data-reduction scheme was developed for this survey to handle 
the large number of \ion{H}{i} spectra ($1.5\times10^6$). The individual 
Hanning smoothed and polarization averaged spectra have an rms brightness 
temperature noise of $\sigma$ = 0.12~K. The final data-cubes have an rms noise 
of $\sigma_{\rm rms} \approx$ 0.05~K and an effective angular resolution of 
$\approx$16\arcmin. In this paper we describe the survey parameters, the 
data-reduction and the general distribution of the \ion{H}{i} gas.\\
The Large Magellanic Cloud (\object{LMC}) and the Small Magellanic Cloud 
(\object{SMC}) are associated with huge gaseous features -- the 
\object{Magellanic Bridge}, the \object{Interface Region}, the 
\object{Magellanic Stream}, and the \object{Leading Arm} -- with a total 
\ion{H}{i} mass of $M$(\ion{H}{i}) = 4.87$\cdot$10$^8$~M$_\odot \left[d/55~{\rm kpc}\right]^2$, 
if all \ion{H}{i} gas is at the same distance of 55~kpc.
Approximately two thirds of this \ion{H}{i} gas is located close to the 
Magellanic Clouds (\object{Magellanic Bridge} and \object{Interface Region}), 
and 25\% of the \ion{H}{i} gas is associated with the \object{Magellanic Stream}. 
The \object{Leading Arm} has a four times lower \ion{H}{i} mass than the 
\object{Magellanic Stream}, corresponding to 6\% of the total \ion{H}{i} mass 
of the gaseous features.\\
We have analyzed the velocity field of the Magellanic Clouds and their 
neighborhood introducing a LMC-standard-of-rest frame. The \ion{H}{i} in the 
\object{Magellanic Bridge} shows low velocities relative to the Magellanic 
Clouds suggesting an almost parallel motion, while the gas in the 
\object{Interface Region} has significantly higher relative velocities 
indicating that this gas is leaving the \object{Magellanic Bridge} building up 
a new section of the \object{Magellanic Stream}. The \object{Leading Arm} is 
connected to the \object{Magellanic Bridge} close to an extended arm of the 
\object{LMC}. 
The clouds in the \object{Magellanic Stream} and the \object{Leading Arm} show 
significant differences, both in the column density distribution and in the 
shapes of the line profiles. The \ion{H}{i} gas in the \object{Magellanic Stream} 
is more smoothly distributed than the gas in the \object{Leading Arm}. 
These morphological differences can be explained if the \object{Leading Arm} is 
at considerably lower z-heights and embedded in a higher pressure ambient 
medium.
               }

   \maketitle
%

\section{Introduction}

The Magellanic Clouds are irregular dwarf galaxies orbiting the Milky Way.
The distances for the Large Magellanic Cloud (\object{LMC}) and the Small 
Magellanic Cloud (\object{SMC}) are 50 and 60~kpc, respectively (see Walker 
\cite{walker} and references therein).
The Magellanic Clouds possess a huge amount of gas -- in contrast to the 
Sagittarius dwarf spheroidal galaxy (\object{Sgr dSph}, Ibata et al. \cite{ibata}), 
which does not currently contain neutral hydrogen (Koribalski et al. 
\cite{koribalski}; Burton \& Lockman \cite{burton}; Putman et al. \cite{putman3}).

The Magellanic Clouds were first detected in \ion{H}{i} by Kerr et al. (\cite{kerr}).
The detailed distribution of \ion{H}{i} in the \object{LMC} was first described 
by Kerr \& de Vaucouleurs (\cite{kerr2}) and McGee (\cite{mcgee}). 
Hindman et al. (\cite{hindman}) discovered that the two galaxies are embedded 
in a common envelope of \ion{H}{i} gas. The gas connecting the Magellanic 
Clouds is called the \object{Magellanic Bridge}. 
The \ion{H}{i} distribution and dynamics in the \object{LMC} were reexamined 
by Rohlfs et al. (\cite{rohlfs}), by Luks \& Rohlfs (\cite{luks}), and more 
recently by Staveley-Smith et al. (\cite{stavele2}). 
Single-dish observations of the \ion{H}{i} gas, using the 64-m Parkes telescope, 
have a spatial resolution of about 225 pc $\left[d/55\,{\rm kpc}\right]$. 
These data have sufficient angular resolution to study the large-scale 
distribution of the \ion{H}{i} gas in the Magellanic System. \ion{H}{i} studies 
at high angular resolution were performed by Kim et al. (\cite{kim}, \cite{kim2}) 
for the \object{LMC} and by Stanimirovic et al. (\cite{stanimirovic},
\cite{stanimirovic3}) for the \object{SMC}, using the Australia Telescope Compact 
Array (ATCA). These observations revealed a fractal structure of the ISM in 
these galaxies and detected structure down to the resolution limit. 

Dieter (\cite{dieter}) surveyed the Galactic poles with the Harvard 60-ft
antenna and noted several detections with high radial velocities near the 
southern Galactic Pole, which have no counterparts towards the northern pole.  
This was the first detection in \ion{H}{i} 21-cm line emission of what we 
today call the ``Magellanic Stream'', but she did not identify it as a coherent 
structure.  Further observations of the northern part of the 
\object{Magellanic Stream}, called the ``south pole complex'' at this time, 
were performed by Kuilenburg (\cite{kuilenburg}) and Wannier \& Wrixon 
(\cite{wannier}). Mathewson et al. (\cite{mathewson}) observed the southern 
part of the \object{Magellanic Stream} using the 18-m Parkes telescope. They 
discovered the connection between the ``south pole complex'' and the Magellanic 
Clouds and called it the ``\object{Magellanic Stream}''.  Later on, several 
parts of the \object{Magellanic Stream} were observed with higher resolution 
and sensitivity (e.g. Haynes \cite{haynes}; Cohen \cite{cohen}; Morras 
\cite{morras},\cite{morras2}; Wayte \cite{wayte}; Stanimirovic et al. 
\cite{stanimirovic2}; Putman et al. \cite{putman2}).

Wannier et al. (\cite{wannier2}) discovered high-positive-velocity clouds 
close to the Galactic Plane. Several clouds in this region were subsequently 
observed with higher resolution and sensitivity (e.g. Mathewson et al. 
\cite{mathewson}; Giovanelli \& Haynes \cite{giovanelli}; Mathewson et al. 
\cite{mathewson3}; Morras \cite{morrasa}; Morras \& Bajaja \cite{morrasbajaja}; 
Bajaja et al. \cite{bajaja}; Cavarischia \& Morras \cite{cavarischia}; Putman 
et al. \cite{putman}; Wakker et al. \cite{wakker2}).

Recently, two large-scale surveys have been completed, covering the southern sky. 
The first is the Argentinian \ion{H}{i} Southern Sky Survey (HISSS), presented 
by Arnal et al. (\cite{arnal}) and Bajaja et al. (\cite{bajaja04}). HISSS is the 
counterpart to the Leiden/Dwingeloo survey of Galactic neutral atomic hydrogen 
(Hartmann \& Burton \cite{hartmann}) and offers an angular resolution of 
30\arcmin\ with grid spacings of 30\arcmin. The velocity resolution is 
$\Delta v$ = 1.3~km\,s$^{-1}$ and the rms noise is $\sigma_{\rm rms} \approx $ 
0.07~K. A combined all-sky \ion{H}{i} survey is presented by Kalberla et al. 
(\cite{kalberla04}).
The second survey is the \ion{H}{i} Parkes All Sky Survey (HIPASS, Barnes et 
al. \cite{barnes}), that was performed using the multi-beam facility of the 
Parkes telescope (Staveley-Smith et al. \cite{stavele1}). 
It offers a velocity resolution of $\Delta v$ = 16~km\,s$^{-1}$ and an rms noise 
of $\sigma_{\rm rms} \approx $ 0.01~K. The observing mode of the HIPASS is 
in-scan beam-switching. This mode filters out all large-scale structure of the 
Milky Way or the Magellanic System. Recently, routines were developed to 
recover the majority of the extended structure in the features of the 
Magellanic System (Putman et al. \cite{putman2}).

This paper presents the Parkes narrow-band \ion{H}{i} survey of the Magellanic 
System, designed not only to study the system at high velocity resolution, but to 
retain sensitivity to all spatial scales. 
Section~2 describes the survey parameters, the observations and the data-reduction 
in detail. Section~3 presents the \ion{H}{i} in the Magellanic Clouds
and their immediate vicinity. Section~4 describes the \object{Interface Region} and 
isolated clouds in the vicinity of the Magellanic Clouds. Sections~5 and 6 present 
the \ion{H}{i} in the \object{Magellanic Stream} and the \object{Leading Arm}, respectively. 
Section~7 gives a summary and conclusions.

\begin{figure*}[t]
\includegraphics[width=8.1cm]{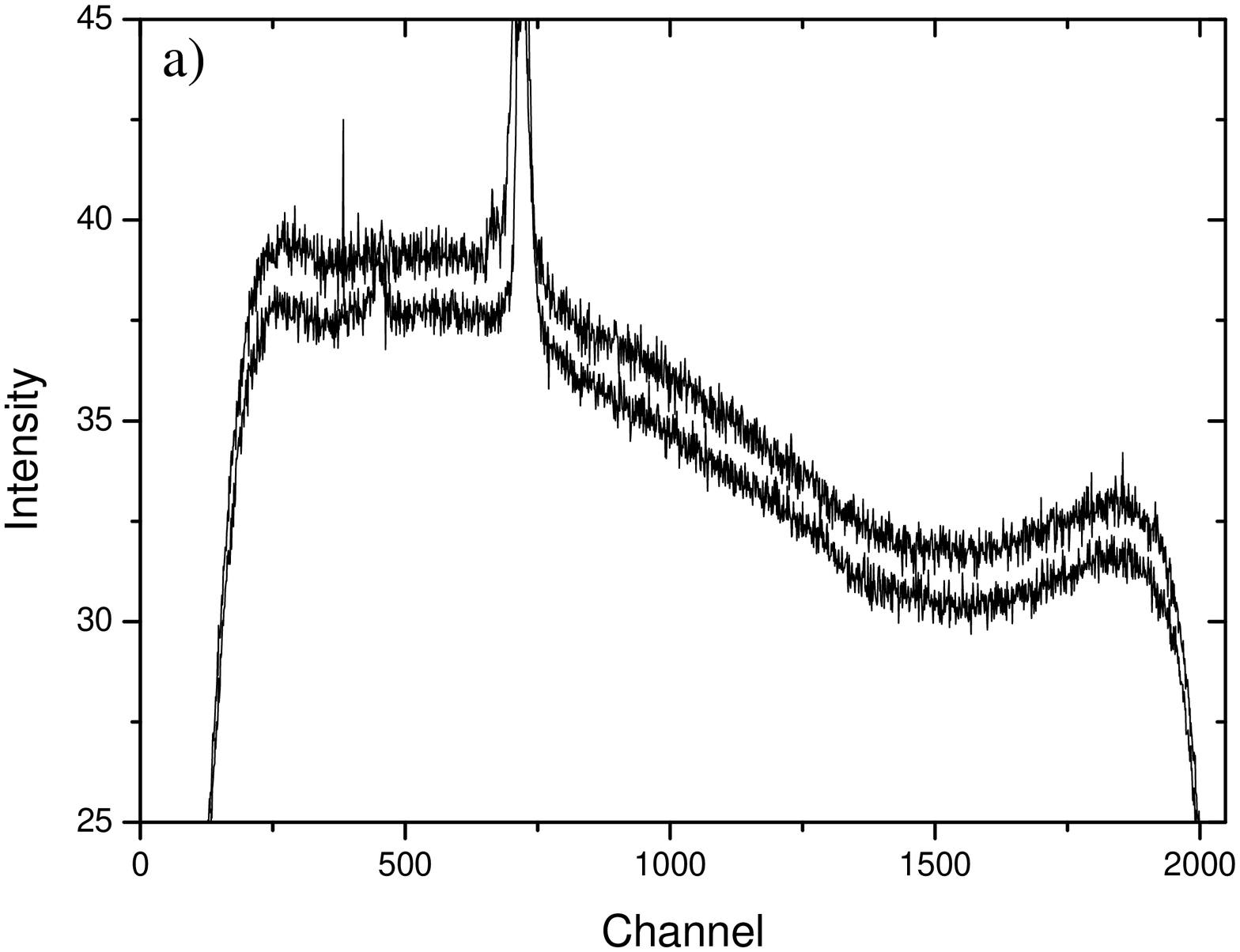}
\includegraphics[width=8.1cm]{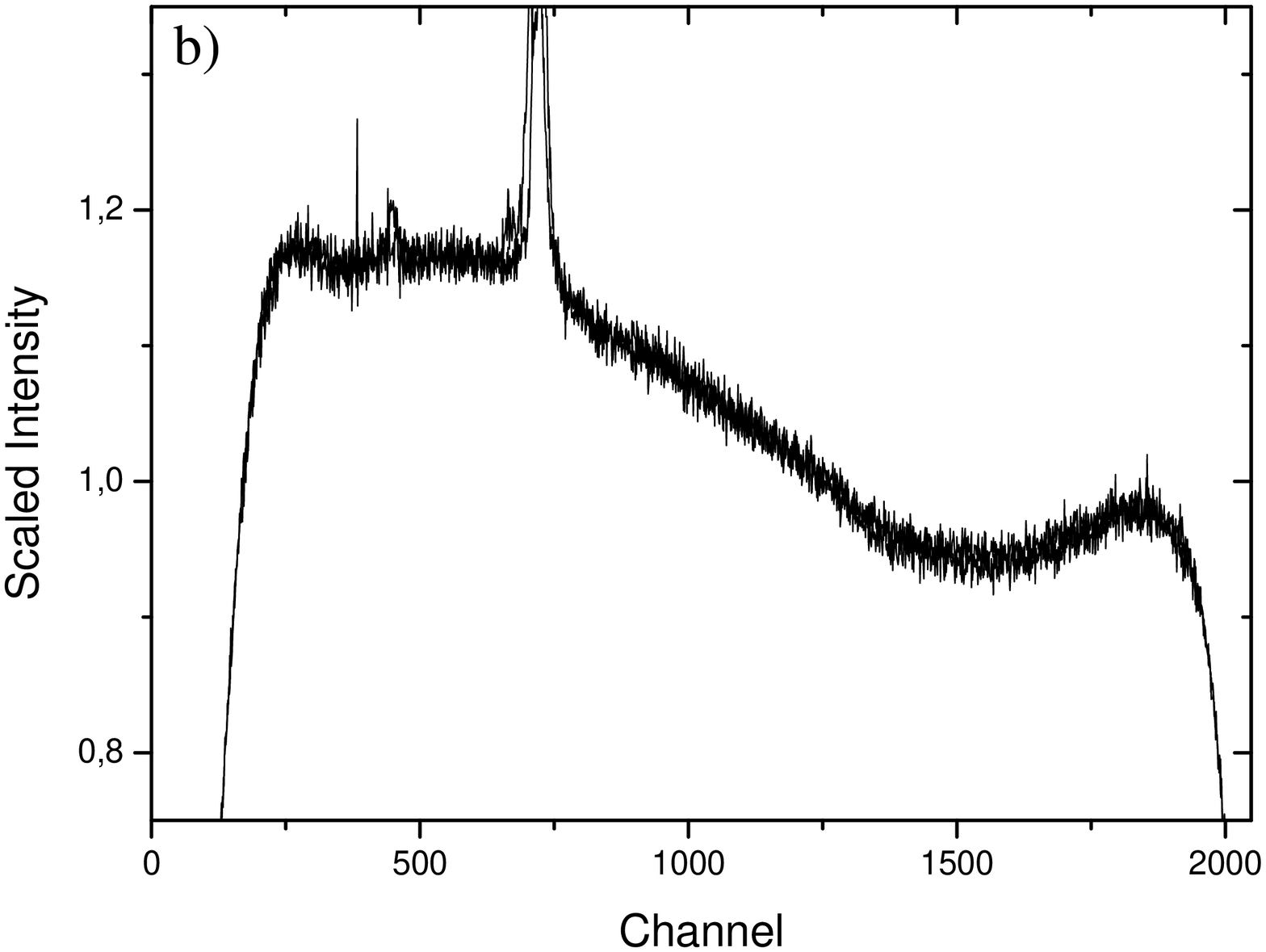}
 \caption{Two example spectra from one scan. {\bf a)} original raw
 spectra. The individual observed bandpasses comprise not only different amounts of 
astronomical line emission, but also a varying amount of continuum emission due to 
astronomical sources, radiation from the ground etc.
{\bf b)} the same spectra after scaling with T$_{\rm sys}$.
The scaled spectra are almost identical, indicating that the shape of the bandpass 
is very stable with time.}
 \label{bandscale}
\end{figure*}
\section{Observations and Data-Reduction}

\subsection{Observations}

The Parkes telescope has a parabolic reflector with a diameter of 64\,m that can be 
operated above  30\degr\ elevation.
At 21\,cm, a multi-beam facility with 13 beams is available, offering a half 
power beam width (HPBW) of 14\farcm1. All 13 beams are, however,
usable only in wide-band mode, operating with a bandwidth of 
64\,MHz with 1024 channels per polarization. 
This mode is well suited for extra-galactic surveys like HIPASS. Observing modes with 
a higher velocity resolution (the narrow-band modes) utilize the central seven beams in 
a hexagonal configuration. These beams are much more symmetric than the outer beams, 
8 -- 13, which suffer from coma distortion. For this survey, we used the narrow-band mode 
with 8 MHz bandwidth, offering 2048 autocorrelator channels for each beam and polarization. 
The bandwidth of 8 MHz corresponds to a velocity coverage of 1690~km\,s$^{-1}$. 
The channel spacing of 3.91 kHz corresponds to a velocity spacing of 0.825~km\,s$^{-1}$, 
while the effective velocity resolution is 1.0~km\,s$^{-1}$. Our survey has therefore a 
16 times higher velocity resolution than HIPASS.

The \ion{H}{i} observations of the Magellanic System were performed within 11 
days from February 17th to 21st and from November 2nd to 8th~1999.
The data were taken using the so-called in-band frequency-switching mode.
The central frequencies of the two spectra (``signal-'' and
``reference-spectrum'') are chosen in a way that the ``reference-spectrum'' also 
contains the spectral line. A frequency offset of 3.5 MHz between both signals 
was chosen for our survey. The advantage of this method is that the telescope is 
on-source all the time.

The multi-beam receiver can be rotated by feed-angles between --60 and 60
degrees. The feed-angle was set to 19\fdg1 relative to the direction of 
the scan to provide uniform sampling. The \ion{H}{i} data were observed in 
on-the-fly mode with a telescope scanning speed of 1\arcmin\ per second. 
The integration time per spectrum was set to 5 seconds for both frequencies. 
The beam moves 5\arcmin\ on the sky during the integration, leading to a slightly 
elongated beam in direction of the scan. The scans were aligned to Magellanic 
coordinates\footnote{The Magellanic coordinates used in this paper are 
{\em not} the coordinates defined by Wannier \& Wrixon (\cite{wannier}). 
The Magellanic coordinate system  was defined by Wakker (\cite{wakker}) as follows: 
the line of zero latitude is defined as the line from Galactic longitude 
$l$~=~90\degr\ across the southern Galactic pole to $l$~=~270\degr. 
Zero longitude is defined by the position of the LMC.}, as both the southern 
poles of Equatorial and Galactic coordinates fall into the region of interest. 
The scans were taken in direction of Magellanic latitude at constant longitude. 
The scans had a length of 15\degr\ for most observed fields. 
The offset between two adjacent scans (in Magellanic longitude) was set to 
66\farcm7. A feed-angle of 19\fdg1 yields beams moving along 
equidistant lines with a separation of 9\farcm358. Each field was 
observed a second time with an offset of 33\farcm6. The combined data have 
grid-spacings of 5\arcmin\ in the direction of the scan (Magellanic Latitude) 
and 4\farcm679 perpendicular to the scan direction (Magellanic Longitude).
The comparison  with the HPBW of 14\farcm1 demonstrates that the data have 
a higher sampling than required by the Nyquist-sampling theorem.

We utilized early HIPASS maps from Putman et al. (\cite{putman2}) to define the 
regions of interest with emission from the Magellanic Clouds and their gaseous arms. 
In total, 745\,598 individual pointings, covering approximately 24 percent of the 
southern sky, were observed.

\subsection{Data-Reduction}

A total number of 1\,491\,196 source spectra were observed for the Parkes 
narrow-band survey of the Magellanic System. Due to the large number of spectra 
an automated reduction pipeline was a necessity. 

\subsubsection{Bandpass Calibration}

We used the standard method to perform the bandpass calibration:
\begin{eqnarray}\label{bandcal}
T_{\rm A} &=& T_{\rm sys}~ \frac{T_{\rm sig} - T_{\rm ref}}{T_{\rm ref}}.
\end{eqnarray}
For in-band frequency switching, both $T_{\rm sig}$ and $T_{\rm ref}$ contain the
spectral lines shifted by 3.5 MHz. In the course of the data-reduction it 
became clear that the baselines are highly non-linear -- a polynomial of 
8th order is needed to fit most baselines. It is extremely difficult to separate 
baseline wiggles from faint emission lines. Even a large amount of work 
failed to produce sufficiently flat baselines with an automated pipeline. 
Accordingly, a different approach was taken which involved searching for individual 
reference positions for each spectral channel.

The individual observed bandpasses not only have different amounts of astronomical 
line emission along the scan, but also a varying amount of continuum emission. 
Figure~\ref{bandscale}a shows two randomly chosen raw-spectra of a scan containing a 
different amount of continuum emission. Figure~\ref{bandscale}b shows the same spectra 
after division by T$_{\rm sys}$. The scaled spectra are almost identical (expect 
for varying line emission).
The spectra of a scan of the same beam, polarization and central frequency
have almost the same shape, while the shape of the bandpass varies strongly 
for different beams, polarizations and central frequencies.

The stability of the shape of the bandpass can be used to perform the bandpass
calibration. Scans were taken in a manner such that each scan has, at a given 
velocity, an appreciable length free of Magellanic emission.
A reference spectrum is compiled by searching for emission free regions for 
each velocity. This is done by dividing the scan into smaller parts (e.g. ten
subscans). For each subscan, beam, polarization, and frequency a median of the 
values for each individual spectral channel is calculated. The minimum of these 
medians for a given channel is taken for each beam, polarization, and frequency. 
The method is similar to the MINMED5 method described by Putman et al. 
(\cite{putman2}).

This method produces 28 bandpasses, one for each beam, polarization and center 
frequency. These bandpasses are free from emission from the Magellanic System, 
but they still contain emission from the extended gas of the Milky Way and 
``extended'' radio frequency interference (RFI) that is apparent in all spectra 
of a scan. Some more steps are necessary to recover also the extended emission, 
e.g. the \ion{H}{i} from the local gas of the Milky Way.
The remaining spectral features have in general line widths of a few channels in 
the case of RFI and 10 to 25 channels in the case of Galactic emission, while the 
baseline varies on much larger numbers of channels with lower amplitude. The bandpass 
for each beam, polarization and center frequency is now searched for lines by looking 
for a steep rise of more than 1~K over 8 channels. These identified regions are enlarged 
by $\pm$50 channels to guarantee that the whole line including possible extended 
line wings are included. An initial 6th-order polynomial is subtracted from the 
bandpasses. Every channel that was classified as being free of 
emission is now re-examined. If the intensity of the fitted spectrum at a channel 
is higher than 0.65~K the channel is classified as having line 
emission.  The newly identified line windows are extended as described above 
and applied to the original bandpasses. The procedure is repeated with a
threshold of 0.3~K. 

A polynomial of 6th order is fitted to the bandpass using the final spectral line mask. 
The values within the line windows are replaced by the polynomial, while the other 
channels remain unchanged. This routine yields 28 reference bandpasses that are free 
from line emission and strong RFI. These final bandpasses are now applied to all spectra
of the scan using Eq.~\ref{bandcal}. 

\subsubsection{Baseline Correction and RFI Subtraction}

The bandpass calibrated spectra have in general quite flat baselines. For most 
spectra ($\approx$ 97\%) a linear baseline subtraction is sufficient. The remaining 
spectra have baselines where a polynomial of third order is needed. To compute the
final baseline, a spectral mask is again required. This mask is calculated as
described above.

The spectra available at this point are bandpass calibrated and have flat
baselines, but are still contaminated by RFI. In the case of the Parkes
narrow-band survey, only sinc-shaped RFI is present. After Hanning smoothing,
spikes with a width of one or two channels are present. These interferences
appear at the same channels over the whole scan (no Doppler tracking has
been applied so far). The identification process averages and Hanning smoothes 
all spectra for each beam, polarization and center frequency. 
Every spectral feature in the averaged spectra with a width of one or two channels
with an intensity larger than 0.25~K is recognized as RFI.
The values at these channels are substituted by a linear interpolation of the
neighboring channels for all individual spectra. 

In the very last step the rest-frame of the velocity is converted from topocentric 
to the local-standard-of-rest (LSR) frame.

\subsubsection{Gain Calibration}

The \ion{H}{i} data were calibrated to brightness temperature scale using the 
standard calibration sources S8 and S9 (Williams \cite{williams}). We used
the well defined flux of S8 at 14\arcmin\ resolution from Kalberla et al. (\cite{kalberla})
as primary calibrator and derived a brightness temperature of $T_{\rm B}$ = 83~K for S9.
The calibration source S9 served as secondary calibrator to monitor the calibration.
The parameters of the calibration sources are summarized in Table \ref{poss8s9}.  

\begin{table}[t]
\caption{Parameters of the calibration sources S8 and S9.}
\begin{tabular}{lccccc}
\hline
 & l & b & RA (J2000) & DEC (J2000) & T$_{\rm max}$ \\
\hline
S8 & 207\fdg00 & --15\fdg00& 5$^{\rm h}$47$^{\rm m}$21.3$^{\rm s}$ & --1${^\circ}$40'18.4" 
& 76~K \\
S9 & 356\fdg00 & --4\fdg00& 17$^{\rm h}$52$^{\rm m}$05.4$^{\rm s}$ & --34${^\circ}$25'15.4" 
& 83~K \\
\hline
\end{tabular}
\label{poss8s9}
\end{table}

The calibrators were observed by each of the seven beams independently and in turn.
The calibration source S8 was observed 4 times and S9 7 times. 
The standard deviations of the calibration factors for each individual beam,
polarization and center frequency are lower than one percent, demonstrating 
that the calibration factors are very stable over a period of a couple of days. 

In addition to the pointed observations, S9 was observed five times using the 
on-the-fly mode to make sure that the results derived from pointed observations 
do not differ from the on-the-fly data. The standard deviation of the intensity
at the position of S9 is 0.53 percent around the nominal value of 83~K.

\subsection{Stray-Radiation}

\begin{table}[t]
\caption{Observational parameters of the \ion{H}{i} survey.}
\begin{tabular}{lc}
\hline
 Antenna diameter & 64 m\\
 HPBW & 14\farcm1\\
 Observing mode & on-the-fly mapping\\
 Sky coverage & $\approx$24\%\\
 Number of spectra & 1\,491\,196\\
 Grid in Magellanic Coord. & 4\farcm679, 5\arcmin\\
 Integration time per spectrum & 5 sec.\\
 Total bandwidth & 8 MHz\\
 Usable bandwidth & 4.5 MHz \\
 Usable velocity coverage & 950 km\,s$^{-1}$\\
 Original velocity resolution & 1.0 km\,s$^{-1}$\\
 Channel separation & 0.825 km\,s$^{-1}$\\
 Velocity resolution after Hanning smooth & 1.65 km\,s$^{-1}$\\
 RMS noise after Hanning smooth & 0.12 K\\
\hline
 {\bf Final data cubes:} & \\
 Angular resolution & 16\arcmin \\
 RMS noise & 0.05 K\\
\hline
\end{tabular}
\label{parameter}
\end{table}

The observed antenna temperature, $T_{\rm A}$, can be transformed to brightness temperature, 
$T_{\rm B}$, using the equation from Kalberla et al. (\cite{kalberla2}):
\begin{eqnarray}\label{strayrad}
T_{\rm B}\left(\vartheta,\phi,v\right) & = & \frac{1}{\eta_{\rm MB}} T_{\rm A}\left(\vartheta,\phi,v\right)\\ \nonumber 
 & & - \frac{1}{\eta_{\rm MB}} \int_{\rm SL} P\left(\vartheta - \vartheta^\prime,\phi - \phi^\prime\right) T_{\rm B}\left(\vartheta^\prime, \phi^\prime,v\right) {\rm d}\Omega^\prime,
\end{eqnarray}
where $\eta_{\rm MB}$ is the mainbeam efficiency and $P$ is the antenna diagram. The second 
line gives the contribution from the side-lobes (SL) also known as stray-radiation. The 
entire sky is covered by \ion{H}{i} emission at low velocities ($v\approx0$\,km\,s$^{-1}$). 
Kalberla et al. (\cite{kalberla2}) demonstrated that the total amount of stray-radiation for 
Galactic emission is typically about 15\% of the observed profile area, but it can increase to 
50\% or more at high Galactic latitudes.

\begin{figure*}[p]
\includegraphics[width=13.8cm]{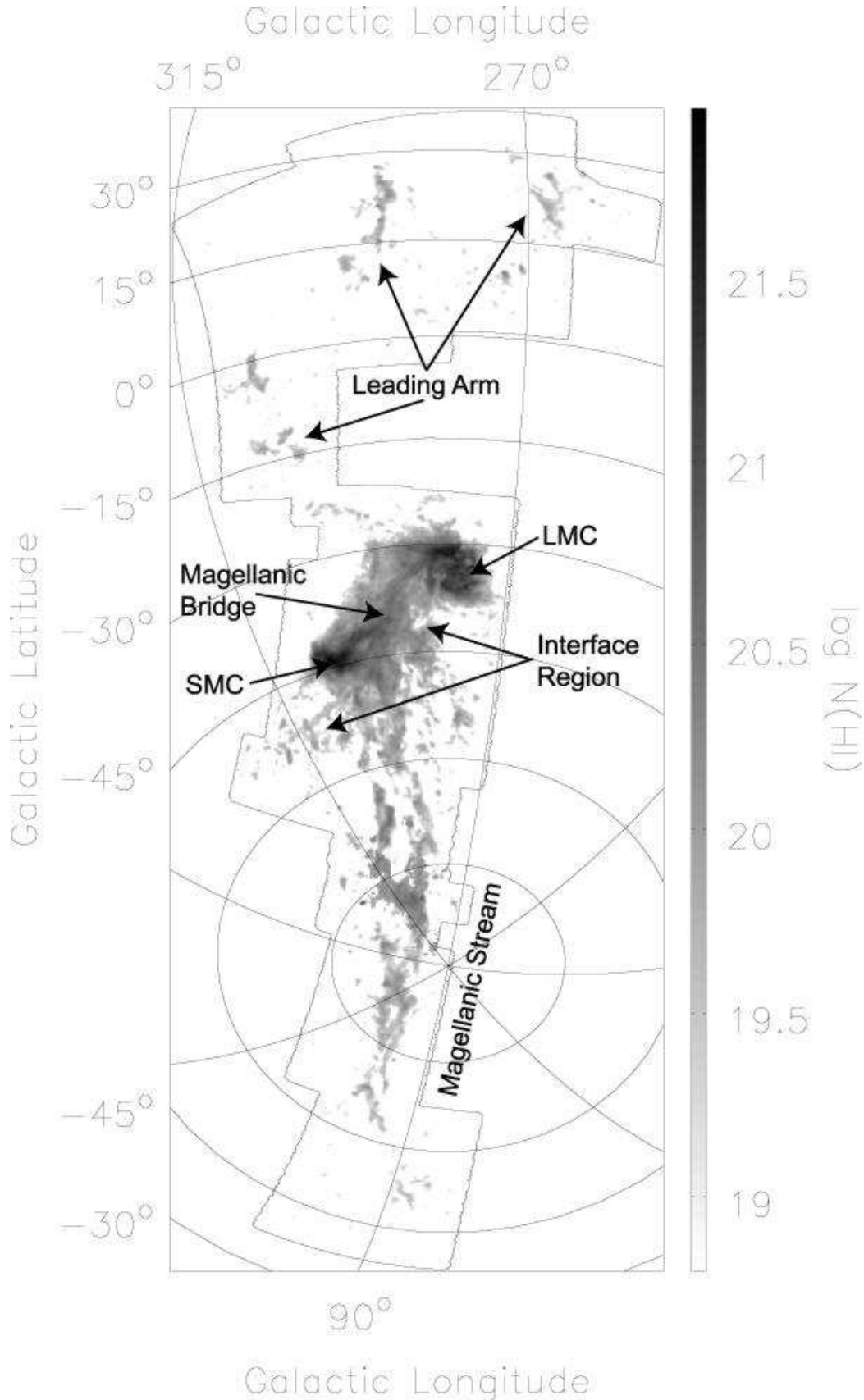}
\caption{\ion{H}{i} column density distribution of the entire Magellanic System. \ion{H}{i} 
emission associated with the Milky Way (see Fig.~\ref{allmagseite}) was omitted. Note that 
the radial velocities of the \object{Magellanic Stream} are close to Galactic velocities 
near ($l$,$b$) = (315\degr, --80\degr). The emission of this gas was recovered as described 
in Sect.~\ref{localgas}. The grey-scale is logarithmic and represents column densities between 
$N$(\ion{H}{i})~=~6$\cdot$10$^{18}$\,cm$^{-2}$ (light grey) and 
$N$(\ion{H}{i})~=~9$\cdot$10$^{21}$\,cm$^{-2}$ (black). The black contours indicate the borders
of the survey. Zero longitude is to the left.}
\label{allmag}
\end{figure*}

\begin{figure*}[t!]
\includegraphics[width=17.5cm]{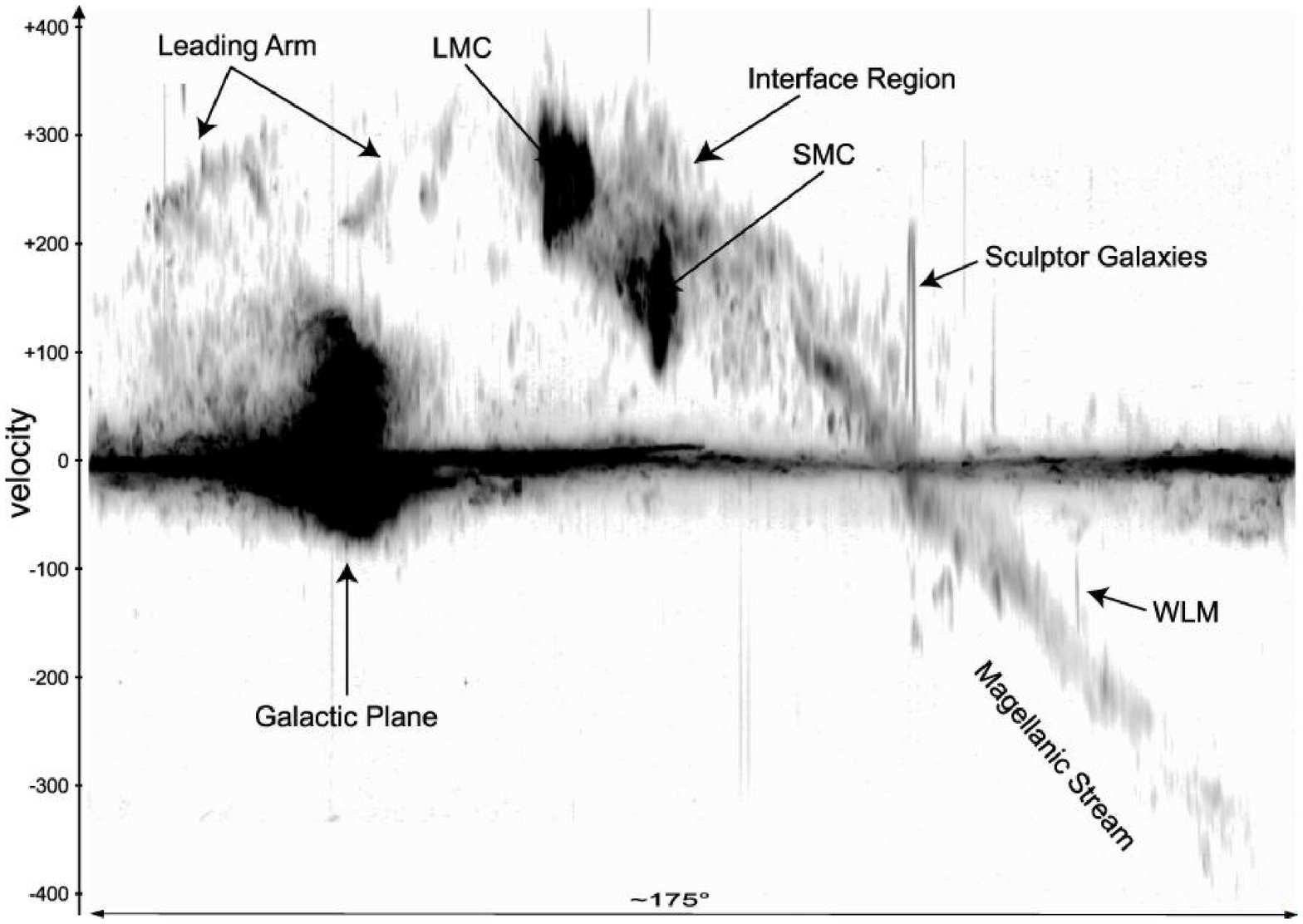}
\caption{The figure shows a position/velocity peak intensity map of the data-cube shown in 
Fig.~\ref{allmag}: radial velocity ($v_{\rm LSR}$) is plotted as a function of position 
(corresponding to the y-axis of the data-cube presented in Fig.~\ref{allmag}). The grey-scale 
indicates the peak intensity  (along the x-axis of the data-cube presented in Fig.~\ref{allmag}). 
White corresponds to $T_{\rm B} = 0$~K, black corresponds to $T_{\rm B} > 15$~K.
The figure shows the Magellanic Clouds and the associated gaseous features, the local gas of 
the Milky Way (near $V_{\rm LSR}$ = 0\,km\,s$^{-1}$), and some galaxies (see Fig.~\ref{streamseite}).}
\label{allmagseite}
\end{figure*}

The Parkes narrow-band \ion{H}{i} survey of the Magellanic System has not been corrected for 
stray-radiation. The Galactic emission is certainly affected by stray-radiation from the near 
and far side-lobes of the antenna pattern. While the general distribution can be studied, no 
quantitative analysis of the Milky Way \ion{H}{i} emission is possible using our data.

However, the neutral hydrogen associated with the Magellanic System is clearly separated from 
Galactic emission, except for a small region close to the southern Galactic Pole. At high 
velocities only a very small fraction of the sky is covered by \ion{H}{i} emission at each 
individual velocity, leading to a negligible integral over the side-lobe region (see Eq.~\ref{strayrad}). 
A stray-radiation correction for the \ion{H}{i} emission of the gaseous features of the Magellanic 
System is therefore not necessary.

\subsection{Data-cubes and final sensitivity}

The \ion{H}{i} data of the Parkes narrow-band survey of the Magellanic System 
have grid-spacings of 5\arcmin\ in the direction of the scan (Magellanic Latitude) 
and 4\farcm679 perpendicular to the scan direction (Magellanic Longitude). At 
each position, two independent spectra (the two polarizations) are available, 
leading to an 1-$\sigma$ rms brightness temperature noise of $\sigma_{\rm rms} \approx$ 0.12~K in 
the averaged and Hanning smoothed spectra.

The aim of this paper is to present the large-scale distribution and morphology 
of the \ion{H}{i} gas in the Magellanic System at high sensitivity. 
We choose Galactic coordinates for all maps in this paper to simplify a comparison 
with features in the Milky Way. A higher signal-to-noise ratio can be achieved by 
modest spacial smoothing. The data-cubes presented in this paper are gridded using 
the {\em GRIDZILLA} tool that is part of {\em AIPS++}. The true angle grid-spacings 
in the data-cubes are set to 4\arcmin. For each grid point, neighboring spectra within 
a diameter of 8\arcmin\ are taken into account. These spectra are weighted by a Gaussian 
with FWHM of 8\arcmin. This spacial smoothing increases the effective beam size to 
$\approx$16\arcmin, but improves the rms noise to about $\sigma_{\rm rms} \approx$ 0.05~K 
at a velocity resolution of $\Delta v$ = 1.65~km\,s$^{-1}$. 

A 1-$\sigma$ signal of $T_{\rm B}$ = 0.05~K corresponds to a column density of
$N$(\ion{H}{i}) = 1.5$\cdot$10$^{17}$\,cm$^{-2}$ per spectral channel 
($\Delta v$ = 1.65~km\,s$^{-1}$). Typical \ion{H}{i} lines from clouds consisting 
of warm neutral gas, however, show line widths of typically 
$\Delta v_{\rm FWHM} \approx$ 25~km\,s$^{-1}$. A Gaussian line with a peak intensity 
at the 5-$\sigma$ level and a line width of $\Delta v_{\rm FWHM} \approx$ 25~km\,s$^{-1}$ 
corresponds to a column density of approximately 
$N$(\ion{H}{i}) = 1.2$\cdot$10$^{19}$\,cm$^{-2}$. These values give an idea of what is 
detectable in the maps we show in this paper, while even weaker lines can be detected
applying further smoothing.

\begin{figure*}[p]
\includegraphics[width=17.0cm]{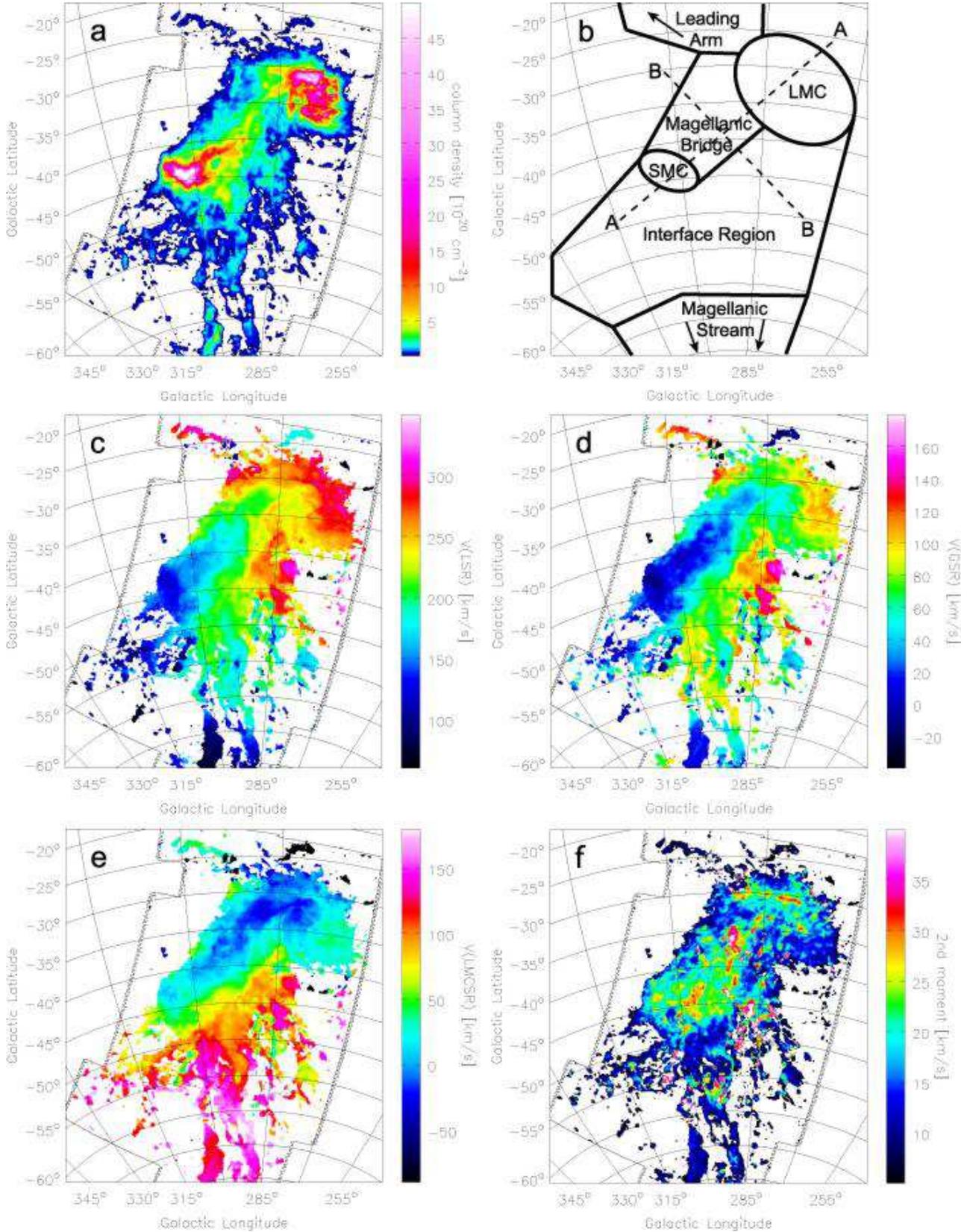}
 \caption{{\bf a)} The \ion{H}{i} column density map of the Magellanic Clouds and their environment. 
 The black contours indicate the borders of the survey. The \object{LMC} and the \object{SMC} are 
 connected by the Magellanic Bridge. The \object{Interface Region} connects the Magellanic Clouds and 
 the \object{Magellanic Bridge} with the \object{Magellanic Stream}.
 {\bf b)} A schematic view of the \object{LMC}, the \object{SMC} and the neighboring \ion{H}{i} gas. 
 The arrows indicate that the \object{Magellanic Stream} and the \object{Leading Arm} continue outside 
 the borders of this map. The dashed lines marked with A and B indicate the position of the slices shown 
 in Figs. \ref{mbslice2} and \ref{mbslice}, respectively. 
 {\bf c)} The velocity field in the local-standard-of-rest frame. 
 {\bf d)} The velocity field in the Galactic-standard-of-rest frame. 
 {\bf e)} The velocity field in the LMC-standard-of-rest frame (see Sect.~\ref{velofield}). 
 {\bf f)} The 2nd moment map.}
 \label{mbvel}
\end{figure*}

The HISSS survey (Arnal et al. \cite{arnal}) has a typical rms of the order 
$\sigma_{\rm rms} \approx$ 0.07~K at a velocity resolution of $\Delta v$ = 1.3~km\,s$^{-1}$. 
HISSS has an angular resolution of 30\arcmin\ and grid spacings of 30\arcmin, i.e. it is not 
fully sampled. Our survey has a higher angular resolution and full sampling. We reach an rms 
noise of the order $\sigma_{\rm rms} \approx$ 0.017~K when smoothing to the angular resolution 
of HISSS.

HIPASS HVC maps have a typical rms of the order $\sigma_{\rm rms} \approx$ 0.008~K at a velocity
resolution of  $\Delta v$ = 26.0~km\,s$^{-1}$ (Putman et al. \cite{putman2}). 
Our survey has an rms noise of $\sigma_{\rm rms} \approx$ 0.012~K when smoothing to the same 
velocity resolution, demonstrating that HIPASS has a slightly higher sensitivity than our survey. 
The higher velocity resolution and the reduction being optimized for the recovery of 
extended emission makes our survey ideal for a detailed study of the Magellanic System.

Figure~\ref{allmag} shows an \ion{H}{i} column density map of the Magellanic Clouds (Sect.~\ref{magclouds}), 
the \object{Magellanic Bridge} (Sect.~\ref{magclouds}), the \object{Interface Region} (Sect.~\ref{sectir}),
the \object{Magellanic Stream} (Sect.~\ref{stream}), and the \object{Leading Arm} (Sect.~\ref{leadarm}).
The map covers the entire area covered by our survey; the contours represent the borders of our survey.
The \ion{H}{i} emission of the Milky Way was omitted. Figure~\ref{allmagseite} shows a position/velocity peak 
intensity map of the data-cube shown in Fig.~\ref{allmag}, where the position corresponds to the y-axis  
of that data-cube. The grey-scale indicates the peak intensity along the x-axis of that data-cube.
The individual gaseous features shown in Figs.~\ref{allmag} and \ref{allmagseite} are discussed in the following sections.


\section{\ion{H}{i} in the Magellanic Clouds and the Magellanic Bridge}\label{magclouds}

\subsection{The column density distribution}\label{mbcolumn}
 
The Large and the Small Magellanic Cloud are \ion{H}{i} rich dwarf irregular galaxies
located on the southern sky at ({\em l,b}) = (280\fdg5, --32\fdg9) and (302\fdg8, --44\fdg3), 
respectively (see Westerlund \cite{westerlund} for a review). 
Figure~\ref{mbvel}a shows the \ion{H}{i} column density distribution of the Magellanic Clouds and 
their environment. Both galaxies are embedded in a common \ion{H}{i} envelope covering tens of 
degrees on the sky. The angular extent of the Magellanic Clouds is therefore not well defined in 
\ion{H}{i}. The gas connecting the Magellanic Clouds is called the \object{Magellanic Bridge} 
(Hindman et al. \cite{hindman}). Figure~\ref{mbvel}b shows a schematic overview of this region and the
location of the individual gaseous features. 

The peak column densities of the \object{LMC} and the \object{SMC} are $N$(\ion{H}{i})~=~5.5$\cdot$10$^{21}$cm$^{-2}$ 
and $N$(\ion{H}{i})~=~1.01$\cdot$10$^{22}$cm$^{-2}$, respectively. \ion{H}{i} column densities above 
$N$(\ion{H}{i})~=~1$\cdot$10$^{21}$cm$^{-2}$ are common in both galaxies. 
Stanimirovic et al. (\cite{stanimirovic}) studied the \ion{H}{i} gas of the \object{SMC} 
in great detail. They analyzed \ion{H}{i} absorption measurements and concluded that regions with \ion{H}{i} 
column densities greater than $N$(\ion{H}{i})~=~2.5$\cdot$10$^{21}$cm$^{-2}$ are affected by 
self-absorption. The peak column densities stated above are therefore only lower limits. 

The \ion{H}{i} gas outside the Magellanic Clouds has much lower column densities. 
Typical column densities in the \object{Magellanic Bridge} are $N$(\ion{H}{i})~$\approx 5\cdot$10$^{20}$cm$^{-2}$. 
This emission can be considered as optically thin (Stanimirovic et al. \cite{stanimirovic}).

The \ion{H}{i} gas of the \object{LMC} has recently been studied in detail by Kim et al. (\cite{kim2}) and by 
Staveley-Smith et al. (\cite{stavele2}). The \object{SMC} was studied recently by Stanimirovic et al. 
(\cite{stanimirovic3}). The \object{SMC} shows an irregular morphology with a high column density tail pointing 
towards the \object{LMC}. A recent study of the \ion{H}{i} in the tail was performed by Muller et al. (\cite{muller2}). 
This tail hosts a stellar component and is also known as Shapleys Wing (Shapley \cite{shapley}). 
The Wing hosts a number of young stars (e.g. Sanduleak \cite{sanduleak}) and molecular clouds 
(Muller et al. \cite{muller1}).
 
\subsection{The Velocity Field}\label{velofield}

The Parkes multi-beam survey of the Magellanic System provides spectra in the
local-standard-of-rest frame. The velocity field of the \ion{H}{i} data is 
analyzed using the first moment of the spectra, i.e. the intensity weighted 
mean velocity, $\overline{v_{\rm LSR}}$, that is calculated by
\begin{equation}
\overline{v_{\rm LSR}} = \frac{\Sigma(T_{{\rm B},i}\cdot v_{i})}{\Sigma T_{{\rm B},i}},
\label{meanvelo}
\end{equation} 
where $T_{{\rm B},i}$ and $v_i$ indicate the brightness temperature and the
radial velocity at channel $i$, respectively.

\begin{figure}[t]
\includegraphics[width=8.8cm]{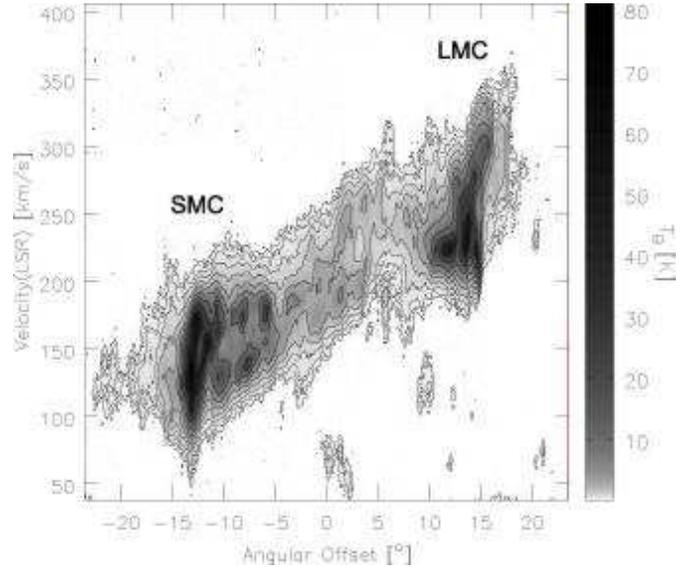}
 \caption{A position/velocity slice through the data-cube shown in Fig.~\ref{mbvel} from 
 ($l$,$b$) = (314\fdg0, -48\fdg0) to (275\fdg0, -25\fdg5). The position of this slice is marked in
 Fig. \ref{mbvel}b by the dashed line A. Contour lines start at $T_{\rm B}$ = 0.15~K (3$\sigma$) 
 and increase by factors of two, i.e. $T_{\rm B}$ = 0.3, 0.6, 1.2, 2.4, 4.8, 9.6, 19.2, 38.4~K.  }
 \label{mbslice2}
\end{figure}

Figure~\ref{mbvel}c shows the observed velocity field of the Magellanic Clouds
and their neighborhood. The observed velocities cover the velocity interval
from $v_{\rm LSR} = $ 120~km\,s$^{-1}$ in the \object{SMC} to $v_{\rm LSR} = $ 300~km\,s$^{-1}$ 
in the \object{LMC}. Figure~\ref{mbslice2} shows an exemplary slice through the data-cube presented 
in Fig.~\ref{mbvel}, demonstrating that the \ion{H}{i} gas  smoothly connects 
both Magellanic Clouds in the position/velocity space. The systemic velocities of the
Magellanic Clouds differ by $\Delta v_{\rm LSR}$~=~125~km\,s$^{-1}$.

The local-standard-of-rest frame is not the best suited reference system as the
Magellanic Clouds form a large system far from the solar neighborhood. 
The varying angle between the line-of-sight and the solar velocity vector
introduces an artificial velocity gradient that is not related to the Magellanic
Clouds. This effect can be overcome by transforming the radial velocities to the
Galactic-standard-of-rest frame using
\begin{equation}
v_{\rm GSR} =  v_{\rm LSR} + 220~{\rm sin}~l~{\rm cos}~b,
\end{equation}
where $l$ and $b$ are Galactic longitude and latitude, respectively. Velocities 
are measured in km\,s$^{-1}$ and 220~km\,s$^{-1}$ corresponds to the velocity 
of the solar circular rotation around the Milky Way. Figure~\ref{mbvel}d shows the 
same region as Fig.~\ref{mbvel}c, but the velocities are transformed to the 
Galactic-standard-of-rest frame. The velocity difference between the Magellanic Clouds 
decreases from $\Delta v_{\rm LSR}$ = 125~km\,s$^{-1}$ to $\Delta v_{\rm GSR}$ = 67~km\,s$^{-1}$. 
The highest radial velocities in the GSR frame are found in the \object{Interface Region}
and the \object{Leading Arm}, while the gas in the \object{Magellanic Bridge} shows low velocities
indicating either low total velocities or velocity vectors that are almost
perpendicular to the lines-of-sight. Unfortunately, the velocity vectors of
\ion{H}{i} clouds are not directly observable. The velocity vectors of the
Magellanic Clouds, however, have been estimated from stellar data recently 
(Kroupa \& Bastian \cite{kroupa}; van der Marel \cite{marel}). The results from 
Kroupa \& Bastian (\cite{kroupa}) demonstrate that the Magellanic Clouds have almost 
parallel velocity vectors, but very different radial velocities due to projection effects. 
We would therefore observe a velocity gradient across the \object{Magellanic Bridge} even if 
all clouds had the same velocity vector. 
The \object{LMC} velocity vector is well determined with relatively low errors and can 
therefore be used as a reference system. The projected radial velocity of the 
\object{LMC} as a function of position of the sky can be expressed by
\begin{equation}
v_{\rm LMC} = v_X~{\rm cos}~l~{\rm cos}~b + v_Y~{\rm sin}~l~{\rm cos}~b + v_Z~{\rm
 sin}~b,
\end{equation}
where ($v_X$, $v_Y$, $v_Z$) = (--56$\pm$39~km\,s$^{-1}$, --219$\pm$23~km\,s$^{-1}$,
186$\pm$35~km\,s$^{-1}$) is the \object{LMC} velocity vector as determined by van der Marel et al.
(2002). The coordinates ($x$,$y$,$z$) are defined in a way, that
the x-axis is in the direction of the Galactic Center, the y-axis in the 
direction of the solar motion, and the z-axis in the direction of the northern 
Galactic Pole. A LMC-standard-of-rest (LMCSR) frame can be defined using the \object{LMC} 
velocity vector as reference. 
\begin{equation}
v_{\rm LMCSR} =  v_{\rm GSR} +56~{\rm cos}~l~{\rm cos}~b +219~{\rm sin}~l~{\rm cos}~b -
186~{\rm sin}~b
\end{equation}
Figure~\ref{mbvel}e shows the velocity field of the Magellanic Clouds and their
neighborhood in the LMCSR frame. The \object{LMC}, the \object{SMC}, the \object{Magellanic Bridge} 
and the first part of the \object{Leading Arm} show very low velocities in the LMCSR frame, 
consistent with gas moving approximately parallel to the Magellanic Clouds.
The hydrogen in the \object{Magellanic Bridge} shows a velocity gradient perpendicular to 
the SMC-LMC axis with almost the same amplitude and orientation as the velocity 
field of the \object{SMC}. This gradient might indicate that this gas has velocity vectors 
comparable to the gas in the \object{SMC}, but the lower gravitational potential outside
the \object{SMC} makes stable orbits unlikely.
In addition, there is a small velocity gradient with decreasing (more negative) 
velocities towards the \object{LMC}, possibly indicating an infall motion of the Bridge 
gas onto the \object{LMC}. 

\begin{figure}[t]
\includegraphics[width=7.8cm]{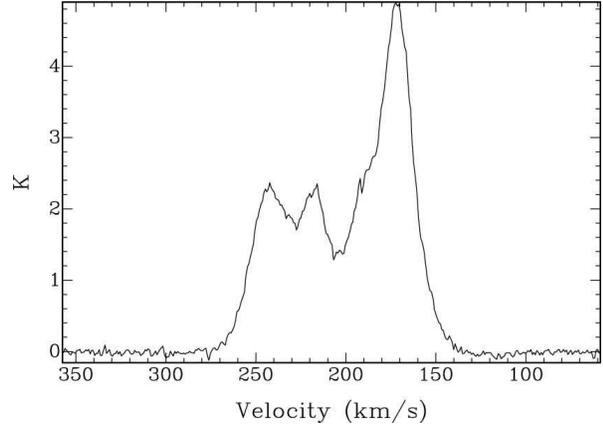}
\caption{An example spectrum ($T_{\rm B}$ vs. $v_{\rm LSR}$) typical for the part 
of the \object{Magellanic Bridge} showing high velocity dispersions that is located at ($l$, $b$) = (290\fdg52,
--39\fdg60). The 2nd moment of this spectrum is $\Delta v = 30.5$~km\,s$^{-1}$, while the 
emission covers a velocity interval of about 125~km\,s$^{-1}$ (see Sect.~\ref{sectdisp}).}
\label{specmb}
\end{figure}

\subsection{Second Moment Analysis}\label{sectdisp}

The 2nd moment of an \ion{H}{i} spectrum indicates the velocity spread of the 
\ion{H}{i} gas within the velocity interval of interest. For a single cloud the 2nd moment 
is an indicator of its internal thermal and turbulent motions. However, within crowded regions 
like the Magellanic Clouds or the \object{Magellanic Bridge}, the 2nd moment can be considered 
as a measure of the relative velocity of individual clouds superposed on the same line-of-sight. 
The 2nd moment of a spectrum is calculated using
\begin{equation}
\Delta v = \sqrt{\frac{\Sigma(T_{{\rm B},i}\cdot(v_{i}-\overline{v})^2)}{\Sigma T_{{\rm B},i}}},
\end{equation}
where $T_{{\rm B},i}$ and $v_i$ indicate the brightness temperature and the
radial velocity at channel $i$, respectively. $\overline{v}$ is the intensity
weighted mean velocity calculated according to Eq.~\ref{meanvelo}.
Figure~\ref{mbvel}f shows the second moment map of the Magellanic Clouds and
their environment. The averaged 2nd moment in the Magellanic Clouds and the
\object{Magellanic Bridge} is $\Delta v \approx 20$~km\,s$^{-1}$ and covers the interval
10~km\,s$^{-1} \le \Delta v \le 35$~km\,s$^{-1}$. 
The values for the \object{Magellanic Bridge} are significantly higher than those of the 
\object{Magellanic Stream} ($\Delta v \approx 15$~km\,s$^{-1}$) and the \object{Leading Arm} 
($\Delta v \approx 8$~km\,s$^{-1}$). There are some 
lines-of-sight in the \object{Interface Region} with extremely high 2nd moments 
($\Delta v \ge 45$~km\,s$^{-1}$), where multiple \ion{H}{i} lines show 
extremely different radial velocities.

These high values for the Bridge are produced by superposed line profiles of clouds at different 
velocities (see Fig.~\ref{specmb}). The individual line profiles show line widths comparable
with those observed in the \object{Magellanic Stream}. The highest values of the 2nd moment
correspond to the regions with the largest number of \ion{H}{i} lines along the line-of-sight. 
They are located along the connecting line of both galaxies in the middle of the Bridge, 
where we also observe the highest column densities. Figure~\ref{mbslice2} shows that multiple
components along the line-of-sight are very common in the \object{Magellanic Bridge}.
Typical relative velocities of the individual line components within the
\object{Magellanic Bridge} are of the order of 50~km\,s$^{-1}$, but some spectra 
show differences of up to 100~km\,s$^{-1}$.

Two clouds on the same line-of-sight having significantly different radial
velocities must have different velocity vectors. Consequently, these two clouds
will disperse if there is no restoring force keeping the clouds together.
In the absence of a restoring force, the relative radial velocity gives an
estimate of the increasing distance between these clouds. A velocity of 50~km\,s$^{-1}$ 
corresponds to a relative distance of 50~kpc after one Gyr. 
For comparison, the current relative distance of the Magellanic Clouds is approximately 
20~kpc. A significant depth along the lines-of-sight through the Magellanic
Bridge is therefore inevitable even if its age is significantly below one Gyr.
However, the long-term stability of the Bridge is not clear, as indicated by the 
large amount of neutral hydrogen escaping towards the \object{Interface Region} (see Sect. \ref{sectir2}).

\subsection{\ion{H}{i} masses}\label{mbmasses}

The Magellanic Clouds are embedded in a common \ion{H}{i} envelope. Therefore,
there is no obvious way to define borders between the \object{LMC}, the \object{SMC}, the Magellanic 
Bridge or the \object{Interface Region}. 
We used both column density and kinematical features to define a subdivision of
the region around the Magellanic Clouds into complexes. Figure~\ref{mbvel}b shows the borders 
that separate the individual complexes. 

The \object{LMC} and the \object{SMC} have within these borders total \ion{H}{i} masses of 
$M$(\ion{H}{i})~=~(4.41$\pm$0.09)$\cdot$10$^8$~M$_{\odot} \left[d/50~{\rm kpc}\right]^2$ and $M$(\ion{H}{i}) = 
(4.02$\pm$0.08)$\cdot$10$^8$ M$_\odot \left[d/60~{\rm kpc}\right]^2$, assuming optically thin emission 
and distances of 50~kpc and 60~kpc for the \object{LMC} and the \object{SMC}, respectively.
The quoted uncertainties are two percent of the derived \ion{H}{i} masses. They include 
residual baseline variations and calibration uncertainties.
A larger error is introduced by the assumption of optically thin emission and the choice of
the borders between the emission from the galaxies and the neighboring gas.

Staveley-Smith et al. (\cite{stavele2}) recently determined an \ion{H}{i} mass of
$M$(\ion{H}{i})~=~(4.8$\pm$0.2)$\cdot$10$^8$~M$_{\odot} \left[d/50~{\rm kpc}\right]^2$ for the \object{LMC}, 
also using the Parkes multi-beam facility. We derive a total \ion{H}{i} mass of 
$M$(\ion{H}{i})~=~(4.6$\pm$0.1)$\cdot$10$^8$~M$_{\odot}$ for the \object{LMC}, if we use the same borders 
between the emission from the \object{LMC} and the neighboring \ion{H}{i} gas as Staveley-Smith et al. 
(\cite{stavele2}). The derived masses from both observations agree within their uncertainties.

Stanimirovic et al. (\cite{stanimirovic}) derived an  \ion{H}{i} mass of 
$M$(\ion{H}{i})~=~(3.8$\pm$0.5)$\cdot$10$^8$~M$_{\odot} \left[d/60~{\rm kpc}\right]^2$ for the \object{SMC}, 
that is in good agreement with our value. Their mass increases to $M$(\ion{H}{i})~=~4.2$\cdot$10$^8$~M$_{\odot}$ 
after applying a self-absorption correction, indicating that the assumption of optically thin emission 
underestimates the true mass by about 10 percent. 

The distances of the \ion{H}{i} clouds outside the two galaxies are not well constrained 
by observations. Numerical simulations suggest that the matter in the \object{Magellanic Bridge} 
is located at similar distances than the Magellanic Clouds (e.g. Yoshizawa \& Noguchi \cite{yoshizawa}). 
Muller et al. (\cite{muller04}) analyzed the spatial power spectrum of the \ion{H}{i} gas 
in the \object{Magellanic Bridge} and concluded that the \object{Magellanic Bridge} consists of two spatially 
and morphologically distinct components. As individual distances for these components are unknown a distance 
needs to be assumed to estimate the \ion{H}{i} mass of the \object{Magellanic Bridge}.
A reasonable estimate for the distance is the average of the distances of the \object{LMC} and the 
\object{SMC}, $d$ = 55\,kpc. 
Using this distance and the borders defined above, the total \ion{H}{i} mass of the \object{Magellanic Bridge} is 
$M$(\ion{H}{i})~=~1.84$\cdot$10$^8$~M$_{\odot} \left[d/55~{\rm kpc}\right]^2$. 
The distance uncertainty dominates all other errors by at least one order of magnitude. The derived 
\ion{H}{i} mass depends on distance quadratically. Increasing the distance estimate from 55~kpc to 
60~kpc increases the derived mass by roughly 20 percent. 


\begin{figure}[t]
\includegraphics[width=8.8cm]{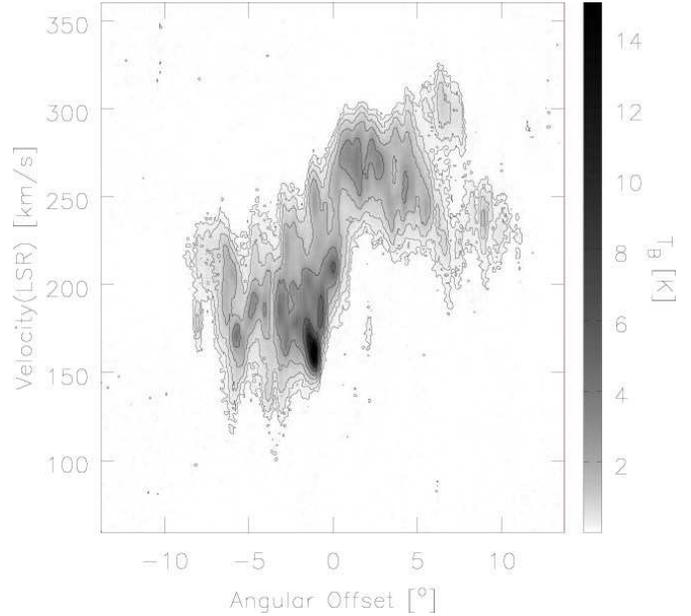}
 \caption{A position/velocity slice through the data-cube shown in Fig.~\ref{mbvel} from 
 ($l$,$b$) = (301\fdg5, -30\fdg5) to (274\fdg0, -50\fdg0). The position of this slice is marked in
 Fig. \ref{mbvel}b by the dashed line B. Negative offset angles indicate 
 emission from the \object{Magellanic Bridge}, positive ones emission from the \object{Interface Region}. 
 Contour lines start at $T_{\rm B}$ = 0.15~K (3$\sigma$) and increase by factors of two, i.e. 
 $T_{\rm B}$ = 0.3, 0.6, 1.2, 2.4, 4.8, and 9.6~K.
 The border between the \object{Magellanic Bridge} and the \object{Interface Region} is defined by the jump in the
 mean velocity at 0\degr\ angular offset.}
 \label{mbslice}
\end{figure}

\section{The Interface Region and Isolated Clouds}\label{sectir}

\subsection{The Interface Region}\label{sectir2}

\begin{figure*}[t!]
\includegraphics[width=18.0cm]{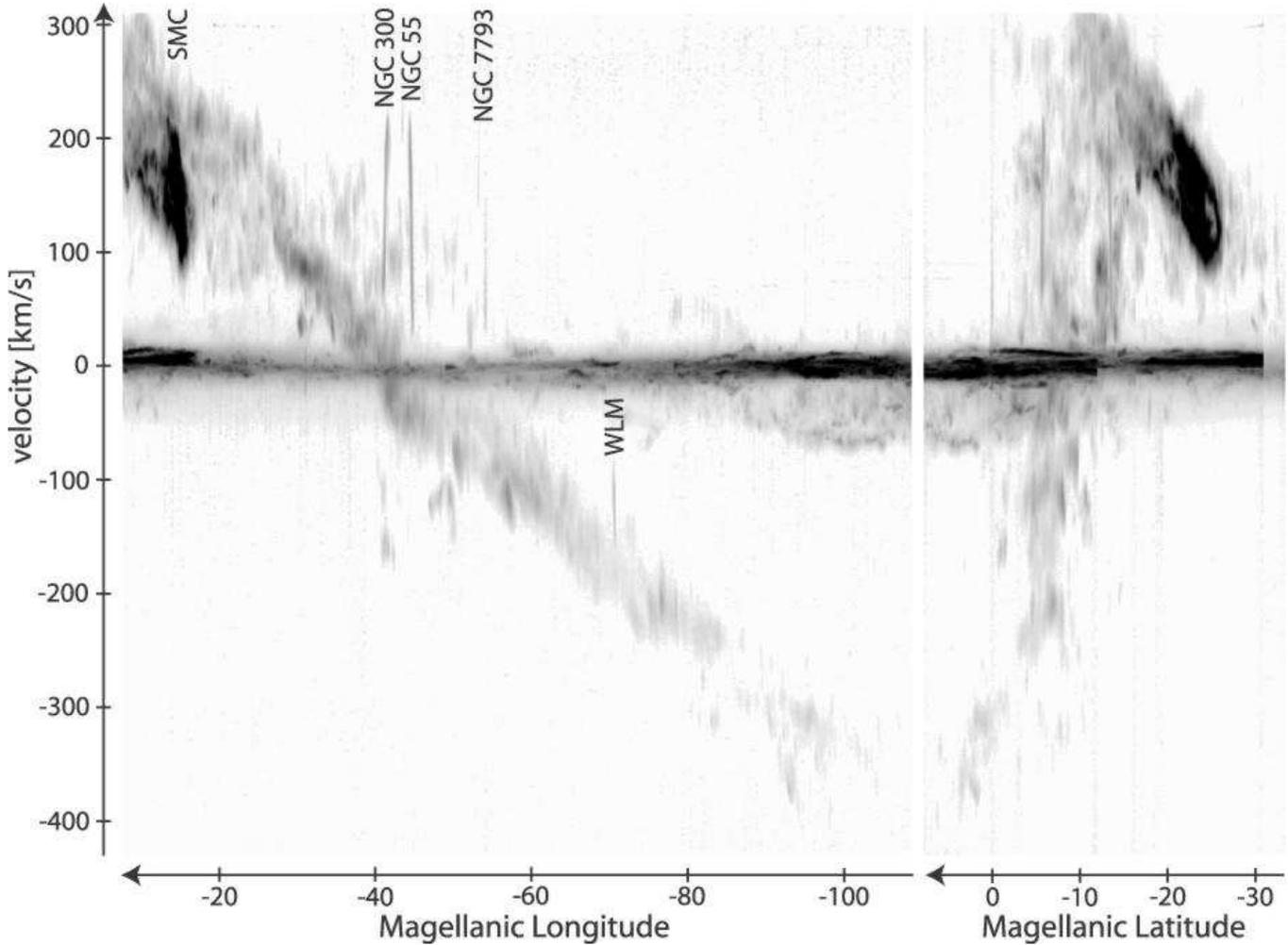}
\caption{The two figures show position/velocity maps of the \object{Magellanic Stream} where the 
grey-scale represents the peak intensity along the third axis. 
{\bf Left:} radial velocity ($v_{\rm LSR}$) is plotted as a function of Magellanic Longitude. The
grey-scale indicates the peak intensity along the Magellanic Latitude axis of the data-cube 
(white corresponds to $T_{\rm B} = 0$~K, black corresponds to $T_{\rm B} > 20$~K). 
{\bf Right:} radial velocity ($v_{\rm LSR}$) is plotted as a function of Magellanic 
Latitude. The grey-scale indicates the peak intensity along the Magellanic 
Longitude axis of the data-cube using the same grey-scale as in the left figure. 
The figures demonstrate that the Magellanic 
Stream forms an almost linear, coherent structure in the position-velocity 
space and covers a velocity interval between +250~km\,s$^{-1}$ and --400~km\,s$^{-1}$.
Moreover, the maps show the emission from the \object{SMC}, the galaxies 
\object{NGC\,300}, \object{NGC\,55}, and \object{NGC\,7793} from the Sculptor Group, and the Local Group 
galaxy \object{WLM}. Some of the vertical lines, e.g. near Magellanic Latitude 0$^\circ$,
are produced by a higher noise close to the border of the mapped region.}
\label{streamseite}
\end{figure*}

The \ion{H}{i} gas close to the \object{SMC} and the \object{Magellanic Bridge} shows a complex filamentary
structure. This area (see Fig. \ref{mbvel}b) will be called \object{Interface Region} henceforward.
Most of these filaments point roughly towards the southern Galactic Pole. The high column density 
filament between Galactic longitude $l$ = 290$^\circ$ and 300$^\circ$ near $b \approx$ --55\degr\ 
is connected to the \object{Magellanic Stream} (Mathewson et al. \cite{mathewson}) that 
continues towards the south passing the southern Galactic Pole (see Sect.~\ref{stream}).

The \ion{H}{i} gas in the \object{Interface Region} has more positive velocities than 
the neighboring gas in the \object{Magellanic Bridge}. This difference in velocity is best visible
in the LMC-standard-of-rest frame (Fig. \ref{mbvel}e). Figure \ref{mbslice} shows a position/velocity 
slice through the data-cube shown in Fig.~\ref{mbvel} from ($l$,$b$) = (301\fdg5, -30\fdg5) 
to (274\fdg0, -50\fdg0). This slice intersects both the \object{Magellanic Bridge} and the 
\object{Interface Region}. The discontinuity in the mean velocity (Fig. \ref{mbslice}) was used to 
define a border between both complexes (compare Figs. \ref{mbvel}b and e). 
The border between the \object{Interface Region} and the \object{Magellanic Stream} can be defined using the 
small gap near ($l$,$b$) = (300\degr, --61\degr).

The distances of the \ion{H}{i} clouds in the \object{Interface Region} are not constrained by observations. 
Numerical simulations suggest distances of the clouds in this region between 30 and 80~kpc 
(e.g. Yoshizawa \& Noguchi \cite{yoshizawa}). As these simulations do not provide a one-to-one mapping 
of the Magellanic System, is it not possible to assign reliable distances to the individual complexes. 
A distance needs to be assumed for a comparison of the \ion{H}{i} masses of the individual complexes. 
A reasonable estimate for this distance is 55~kpc, the average of the distances of the \object{LMC} and the 
\object{SMC}. Using this distance and the borders defined above (see Fig. \ref{mbvel}b), the total \ion{H}{i} 
mass of the \object{Interface Region} is $M$(\ion{H}{i})~=~1.49$\cdot$10$^8$ M$_\odot \left[d/55~{\rm kpc}\right]^2$. 

The huge amount of neutral hydrogen in the \object{Interface Region}, the alignment with the 
\object{Magellanic Bridge} and the smooth connection to the \object{Magellanic Stream}, both in 
position and in radial velocity, indicates that the \object{Interface Region} is closely related to 
the ongoing interaction of the Magellanic Clouds. The high velocity relative to the Magellanic Clouds 
(Fig. \ref{mbvel}e) indicates that this material is currently leaving this region probably building a 
new section of the \object{Magellanic Stream}. The results suggest that the \object{Magellanic Stream} 
was most likely not created as a whole during a past encounter of the Magellanic Clouds, but is still 
in the continuous process of evolution (see also Putman et al. \cite{putman2}). 

Positive radial velocities, both relative to the Galactic-standard-of-rest and to the Magellanic 
Clouds, tend to indicate increasing distances and therefore higher orbits for the clouds in the 
\object{Magellanic Stream}. However, the space motion of the clouds outside the Magellanic Clouds is 
unknown and final conclusions about the three dimensional dynamics can only be made by comparing 
realistic numerical simulations with results from observations.

\subsection{Isolated Clouds Close to the Magellanic Clouds}

The \ion{H}{i} on the northern boundary of the Bridge (in Galactic coordinates) appears much less
structured than the \ion{H}{i} in the \object{Interface Region}, although there are numerous
clouds where the \object{Magellanic Bridge} adjoins the \object{LMC} near ($l$,$b$) $\approx$ 
(290$^\circ$, --27$^\circ$). This filament is part of the \object{Leading Arm} (Putman et al. 1998) 
and continues towards the Galactic Plane (see Sect.~\ref{leadarm}).
The filament has a velocity of $v_{\rm LSR} \approx$~300~km\,s$^{-1}$, comparable to the velocity 
of the \object{Magellanic Bridge} in this region. 

There are also some clouds with lower radial velocities in this region. The largest cloud is
located close to the \object{LMC} at ($l$,$b$) = (283$\degr$, --23\fdg5). It shows an elongated 
structure and a radial velocity of $v_{\rm LSR} \approx$ 180~km\,s$^{-1}$ (see Fig.~\ref{mbvel}). 
In addition, there are two isolated clouds located near ($l$,$b$) = (304\fdg5, --37\fdg3) 
at $v_{\rm LSR} = 110$~km\,s$^{-1}$ that were classified as compact high-velocity clouds (HVCs) by
Putman et al. (\cite{putman4}). 
There are several compact and faint clouds at similar velocities close to the Magellanic Clouds and
also on the same line-of-sight as the \object{LMC} (see Fig. \ref{mbslice2}). 
Some of these clouds have been detected in absorption against \object{LMC} stars (see Bluhm et al. \cite{bluhm} 
and references therein). They must consequently have distances below 50~kpc (the distance of the \object{LMC}), 
indicating that they are foreground objects either of Magellanic origin or ``normal'' HVCs
accidentally located in this region. 
The observed metallicities (Bluhm et al. \cite{bluhm}) and the detection of molecular hydrogen 
(Richter et al. \cite{richterlmc}) are consistent with a Magellanic origin, but a Galactic origin cannot 
be excluded.

The compact and isolated clouds in this region have typical peak column densities within the interval 
2$\cdot$10$^{19}$cm$^{-2} \le N$(\ion{H}{i}) $\le$ 5$\cdot$10$^{19}$cm$^{-2}$. 
The low column densities involve a relatively low total \ion{H}{i} mass associated with these clouds of
$M$(\ion{H}{i})~=~1.2$\cdot$10$^6$ M$_\odot \left[d/55~{\rm kpc}\right]^2$. A distance of 55\,kpc was used 
for the \ion{H}{i} mass to demonstrate that this gas does not affects the mass estimate of the \object{LMC}, 
the \object{SMC}, the \object{Magellanic Bridge}, or the \object{Interface Region}.

\begin{figure}[t]
\includegraphics[width=8.6cm]{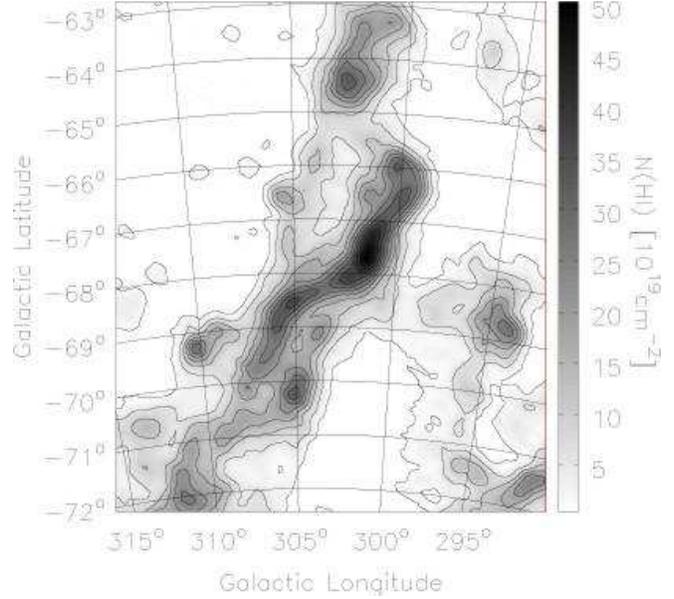}
\caption{\ion{H}{i} column density map of the high column density filament in the positive velocity part 
of the \object{Magellanic Stream}. The contour-lines indicate \ion{H}{i} column densities of 
$N$(\ion{H}{i}) = 1$\cdot$10$^{19}$cm$^{-2}$ and $N$(\ion{H}{i}) = 5$\cdot$10$^{19}$cm$^{-2}$, further 
increasing in steps of $N$(\ion{H}{i}) = 5$\cdot$10$^{19}$cm$^{-2}$.}
\label{ms1}
\end{figure}

\section{\ion{H}{i} gas in the Magellanic Stream}\label{stream}

\subsection{The overall morphology}\label{streamoverall}

The \object{Magellanic Stream} is a $\approx$~100\degr\ long coherent structure that is connected 
to the \object{Interface Region} (see Fig. ~\ref{allmag} for an overview of the column density
distribution). There is no clear starting point of the \object{Magellanic Stream}. We have defined a 
separation between the \object{Interface Region} and the \object{Magellanic Stream} near 
($l$,$b$) = (300\degr, --61\degr).

The \object{Magellanic Stream} appears to be much more confined than the \object{Interface Region}. 
The data clearly show that a simple subdivision in six clouds MS\,I to MS\,VI 
(Mathewson et al. \cite{mathewson4}) is not appropriate with the current resolution and sensitivity. 
Figure~\ref{streamseite} shows a peak intensity map of the \object{Magellanic Stream} in the 
position/velocity space, where radial velocity is plotted on the y-axis, while the 
position is plotted on the x-axis. This figure illustrates the distribution of the neutral hydrogen 
in the \object{Magellanic Stream} relative to Milky Way gas.
The radial velocity changes dramatically over the extent of the \object{Magellanic Stream} 
starting at $v_{\rm LSR} \approx $ +250~km\,s$^{-1}$ near the Magellanic
Bridge and decreasing to $v_{\rm LSR} \approx $ --400~km\,s$^{-1}$ near 
($l$,$b$) = (90\degr, --45\degr), forming an almost linear structure in the position/velocity space. 
This velocity difference of $\Delta v_{\rm LSR}$ = 650~km\,s$^{-1}$ decreases to 
$\Delta v_{\rm GSR}$ = 390~km\,s$^{-1}$ in the Galactic-standard-of-rest frame. 
While the projection effect from the solar velocity vector explains a significant
fraction of the velocity difference, the majority of the velocity gradient is related
to the \object{Magellanic Stream} itself. The observed velocity gradient is inconsistent with a 
circular orbit of the \object{Magellanic Stream}. There are, however, various types of orbits that could
explain the observed velocity gradient.

\begin{figure*}[t]
\includegraphics[width=18.0cm]{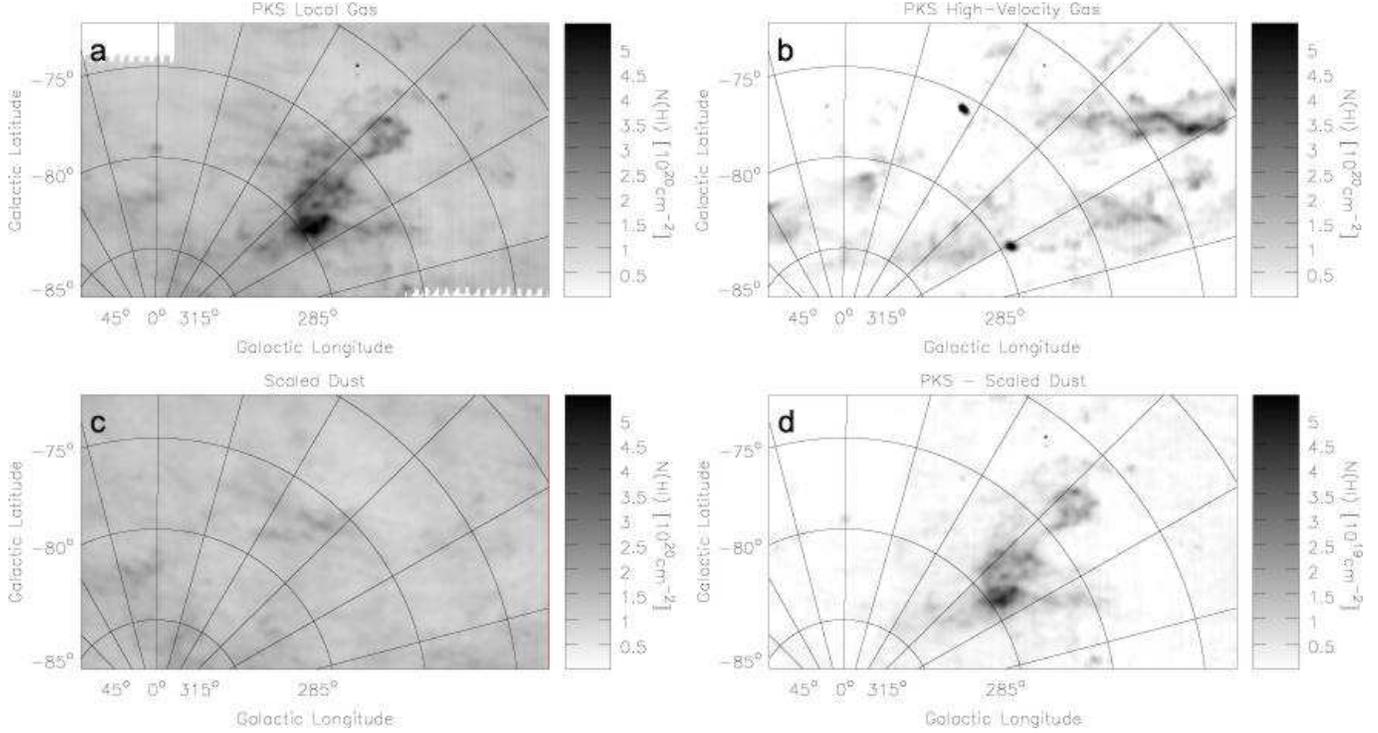}
\caption{The maps show the region where the \object{Magellanic Stream} has radial velocities 
comparable to the local gas of the Milky Way.
 {\bf a)} The \ion{H}{i} column density integrated between $-40$~km\,s$^{-1} \le v_{\rm LSR} \le +30$~km\,s$^{-1}$. 
 {\bf b)} The \ion{H}{i} column density integrated between $-280$~km\,s$^{-1} \le v_{\rm LSR} \le -40$~km\,s$^{-1}$
 and $+30$~km\,s$^{-1} \le v_{\rm LSR} \le +250$~km\,s$^{-1}$. 
 {\bf c)} The dust column density (Schlegel et al. \cite{schlegel}) scaled to \ion{H}{i} 
 column densities. 
 {\bf d)} The difference map subtracting the column densities of map c from those of map a. 
 The residual \ion{H}{i} column density in this region is most likely associated with the 
 \object{Magellanic Stream}.}
 \label{mslocal}
\end{figure*}

The description of the \object{Magellanic Stream} in this section will be subdivided into 
four parts: \ion{H}{i} gas at positive velocities, gas close to Galactic velocities 
and gas at low and high negative velocities.

\subsection{The Magellanic Stream at positive velocities}

The first part of the \object{Magellanic Stream}, also known as MS\,I, has positive radial 
velocities. This part of the \object{Magellanic Stream} comprises two parallel filaments 
and a number of smaller clouds close to the main filaments. The \ion{H}{i}~mass in this part 
of the \object{Magellanic Stream} is $M$(\ion{H}{i}) = 4.3$\times$10$^7$ M$_\odot \left[d/55~{\rm kpc}\right]^2$. 

The filament between ($l$,$b$)~=~(300\degr, --65.5\degr) and ($l$,$b$)~=~(308\degr, --70.5\degr) 
shows the highest column densities in the part of the \object{Magellanic Stream} that is 
not confused by Milky Way gas. The peak column density of $N$(\ion{H}{i}) = 4.7$\cdot$10$^{20}$~cm$^{-2}$ 
is located at ($l$,$b$)~=~(301.5\degr, --67.7\degr) (see Fig.~\ref{ms1}). 
The total \ion{H}{i}~mass of this filament (only the main filament without the neighboring clouds) 
is $M$(\ion{H}{i}) = 1.4$\cdot$10$^7$~M$_\odot \left[d/55~{\rm kpc}\right]^2$. The \ion{H}{i}~mass 
of this filament is comparable to the mass of a very low-mass galaxy and about three times higher 
than the cloud close to M31 discovered by Davies (\cite{davies}). This filament might represent the 
very early stage of a newly born dwarf galaxy. A detailed analysis of this filament will be presented 
in a separate paper.

\subsection{The Magellanic Stream at Galactic velocities}\label{localgas}

\begin{figure}[t]
\includegraphics[width=8.5cm]{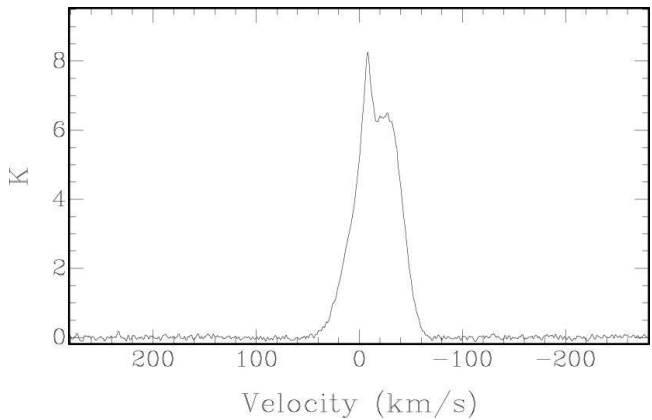}
\caption{A spectrum ($T_{\rm B}$ vs. $v_{\rm LSR}$) of the line-of-sight with the highest 
column density in MS\,II at ($l$,$b$) = (305\fdg5, $-$79\fdg5). The local gas of the Milky 
Way has a velocity of $v_{\rm LSR}$ = $-$8~km\,s$^{-1}$, the \object{Magellanic Stream} has a 
velocity of $v_{\rm LSR}$ = $-$25~km\,s$^{-1}$.}
\label{mslocspec}
\end{figure}

The second part of the \object{Magellanic Stream}, also known as MS\,II, is difficult to 
analyze, because it shows radial velocities comparable to those of the local 
ISM of the Milky Way. 
Fortunately, this part of the \object{Magellanic Stream} is located close to the southern Galactic 
Pole, where the Milky Way shows low column densities.
Figure \ref{mslocal}a shows the \ion{H}{i} column density of this region integrated
between $-40$~km\,s$^{-1} \le v_{\rm LSR} \le +30$~km\,s$^{-1}$. The \ion{H}{i} gas
in this region is relatively smoothly distributed with typical column densities of
$N$(\ion{H}{i}) = 1 to 2$\cdot$10$^{20}$~cm$^{-2}$, but there is a high column density region
near ($l$,$b$) = (305\fdg5, $-$79\fdg5). 
Figure \ref{mslocal}b shows the column density of the \ion{H}{i} gas not confused by Galactic 
emission (integrated over $-280$~km\,s$^{-1} \le v_{\rm LSR} \le -40$~km\,s$^{-1}$
 and $+30$~km\,s$^{-1} \le v_{\rm LSR} \le +250$~km\,s$^{-1}$). The region with high column densities
 is located exactly in that area where the \object{Magellanic Stream} shows velocities comparable to the
 local gas of the Milky Way. Figure \ref{mslocspec} shows a spectrum of the position with the 
 highest column density of $N$(\ion{H}{i}) = 5.9$\cdot$10$^{20}$~cm$^{-2}$ at ($l$,$b$) = 
 (305\fdg5, $-$79\fdg5). The line profiles of the Milky Way and the \object{Magellanic Stream} partly 
 overlap. 

\begin{figure*}[t]
\includegraphics[width=16.5cm]{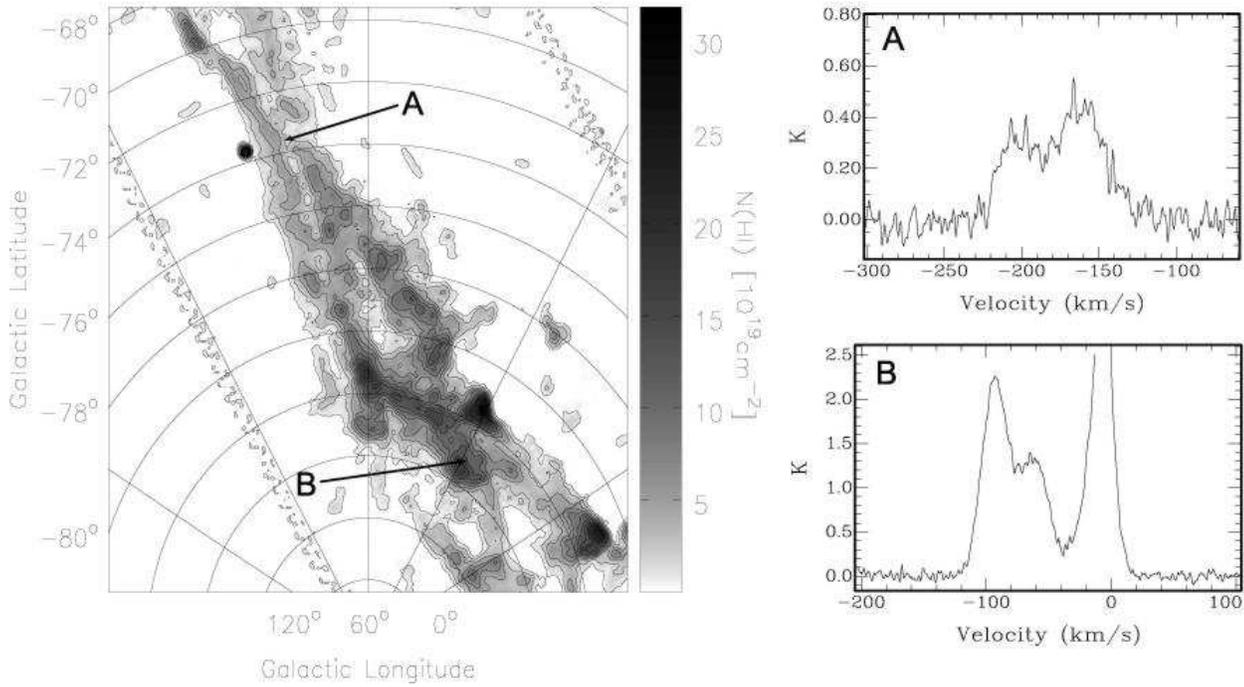}
\caption{\ion{H}{i} column density map of the part of the \object{Magellanic Stream} showing 
negative velocities. The contour-lines indicate \ion{H}{i} column densities of 
$N$(\ion{H}{i}) = 0.5$\cdot$10$^{19}$cm$^{-2}$ and $N$(\ion{H}{i}) = 2$\cdot$10$^{19}$cm$^{-2}$, 
further increasing in steps of $N$(\ion{H}{i}) = 2$\cdot$10$^{19}$cm$^{-2}$.
The two spectra show lines-of-sight where the two filaments of the \object{Magellanic Stream} 
are crossing. Spectrum A is located at ($l$,$b$) = (71\fdg4, --73\fdg5) and spectrum B at 
($l$,$b$) = (29\fdg4, --83\fdg4). The two filaments have different radial velocities differing 
by $\Delta v~=~41$~km\,s$^{-1}$ and $\Delta v~=~31$~km\,s$^{-1}$ for spectra A and B.}
\label{streamnegvel}
\end{figure*}

Fong (\cite{fong}) used IRAS data to demonstrate that there is no detectable dust emission 
associated with the \object{Magellanic Stream}. Schlegel et al. (\cite{schlegel}) produced 
dust column density maps using IRAS and DIRBE data and found that there is a good correlation 
between \ion{H}{i} column density and dust emission in low column density areas like the region 
close to the Galactic Poles. The non-detection of dust emission of the \object{Magellanic Stream} 
can be used to disentangle the column densities of the Milky Way and the Stream. We used the dust 
extinction maps from Schlegel et al. (\cite{schlegel}), smoothed to the same angular resolution as 
the Parkes data, and regions without Magellanic emission to derive a linear 
correlation\footnote{This method can be used to recover emission from the \object{Magellanic Stream} 
as the stray-radiation at Galactic velocities is expected to be quite smooth. However, the numbers 
stated in Eq. \ref{scale} should not be used for quantative analyses.} between the dust extinction 
and \ion{H}{i} column density:
\begin{equation}
N({\rm \ion{H}{i}}) = 6.85\cdot10^{22}~{\rm cm}^{-2}~{\rm mag}^{-1}~E(B-V) + 3.7\cdot10^{20}~{\rm cm}^{-2}.\label{scale}
\end{equation}
Figure \ref{mslocal}c shows the dust emission map scaled to \ion{H}{i} column densities using 
Eq.~\ref{scale}. Figure \ref{mslocal}d shows the residual column density after subtracting the column 
densities from Fig. \ref{mslocal}c from those of Fig. \ref{mslocal}a. The residual column density 
corresponds to \ion{H}{i} gas that is not traced by dust emission and is most likely associated with 
the \object{Magellanic Stream}. 
The regions that appear to be free from Magellanic emission show a 1-$\sigma$ rms noise of 
$N$(\ion{H}{i}) $\approx$ 1.2$\cdot$10$^{19}$~cm$^{-2}$. The map shows a bright peak with a column 
density of $N$(\ion{H}{i}) = 4.8$\cdot$10$^{20}$~cm$^{-2}$ that corresponds to a 40-$\sigma$ detection.
The emission of the \object{Magellanic Stream} has a peak intensity comparable to the local gas of the 
Milky Way, but a larger line width resulting in a larger column density. The total \ion{H}{i}~mass from 
the \object{Magellanic Stream} in this region, $M$(\ion{H}{i}) = 4.3$\cdot$10$^7$~M$_\odot \left[d/55~{\rm kpc}\right]^2$,
is comparable to the \ion{H}{i}~mass of the part of the \object{Magellanic Stream} presented in the 
previous section. Solely emission above the 3-$\sigma$ level was taken into account for the mass estimate. 
The estimated mass is therefore less accurate, because of the residual confusion with Milky Way emission.

\subsection{The Magellanic Stream at negative velocities}

The third part of the \object{Magellanic Stream} (MS\,III to MS\,IV) exhibits increasingly 
negative velocities. Figure~\ref{streamseite} shows peak 
intensity maps in the position-velocity space. This part of the Magellanic 
Stream forms a continuous feature, having much lower intensities than the first 
two parts. The two filaments in this region (see Fig.~\ref{streamnegvel}) have a 
helix-like structure (see also Putman et al. \cite{putman2}). The two spectra from 
Fig.~\ref{streamnegvel} show positions where the two filaments are located on the same line-of-sight.
The two filaments are clearly separated in velocity with radial velocities differing by 
$\Delta v = 41$~km\,s$^{-1}$ (for spectrum A) and  $\Delta v = 31$~km\,s$^{-1}$ (for spectrum B). 
The three dimensional distribution of the gas cannot be derived from \ion{H}{i} data alone.
Nevertheless, the helical structure might indicate the footprint of the Magellanic Clouds, 
while rotating around each other in the past, consistent with a continuous stripping of gas 
from the \object{Magellanic Bridge} as indicated in Sect.~\ref{velofield}. 
The angular width of the filaments is decreasing towards more negative velocities. 
The highest column density of $N$(\ion{H}{i}) = 3.1$\cdot$10$^{20}$~cm$^{-2}$ 
is located at ($l$,$b$)~=~(31.2\degr,  --81.5\degr). This cloud shows more negative velocities 
($\Delta v_{\rm LSR} = 20$~km\,s$^{-1}$) than the neighboring clouds in the main filament of the 
\object{Magellanic Stream}, but it is smoothly connected with the main stream showing a continuous 
velocity gradient. The cloud has an \ion{H}{i}~mass of 
$M$(\ion{H}{i}) = 1.4$\cdot$10$^6$~M$_\odot \left[d/55~{\rm kpc}\right]^2$. 
The total \ion{H}{i}~mass of the third part of the Stream, south of $b$ = --70\degr, is 
$M$(\ion{H}{i}) = 3.0$\cdot$10$^7$~M$_\odot \left[d/55~{\rm kpc}\right]^2$. The clouds in 
this part of the Stream have line-widths of the order of 
$\Delta v_{\rm FWHM}$ = 30 -- 35~km\,s$^{-1}$.

\begin{figure*}[t]
\includegraphics[width=17.5cm]{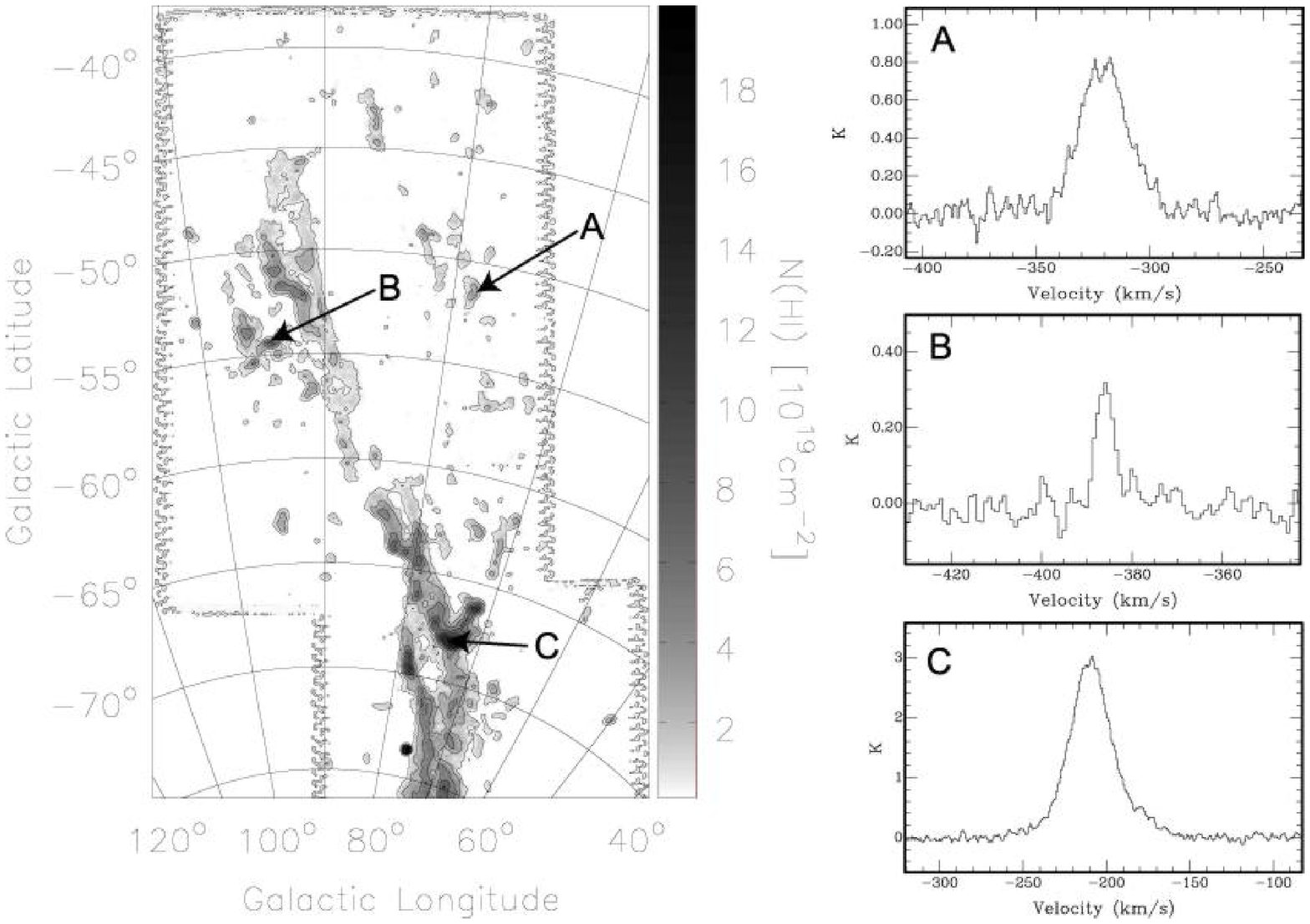}
\caption{\ion{H}{i} column density map of the part of the \object{Magellanic Stream} showing high negative 
velocities. The contour-lines indicate \ion{H}{i} column densities of $N$(\ion{H}{i}) = 0.5$\cdot$10$^{19}$cm$^{-2}$ 
and $N$(\ion{H}{i}) = 2$\cdot$10$^{19}$cm$^{-2}$, further increasing in steps of $N$(\ion{H}{i}) = 2$\cdot$10$^{19}$cm$^{-2}$.
Three exemplary spectra located at ($l$,$b$) = (78\fdg6, --51\fdg6), (93\fdg6, --53\fdg7), and (73\fdg4, --68\fdg0)
are shown to the left. These \ion{H}{i} lines are well represented by single Gaussians with
line widths of $\Delta v_{\rm FWHM}$ = 23~km\,s$^{-1}$, $\Delta v_{\rm FWHM}$ = 3.9~km\,s$^{-1}$, and 
$\Delta v_{\rm FWHM}$ = 28~km\,s$^{-1}$ for spectra A, B, and C, respectively.
}
\label{streamnord}
\end{figure*}

\subsection{The Magellanic Stream at high negative velocities}

The last section of the \object{Magellanic Stream} (MS\,IV to MS\,VI) has very high negative 
radial velocities (up to $v_{\rm LSR}$ = \mbox{--400~km\,s$^{-1}$}) and low column 
densities. Figure~\ref{streamnord} shows the column density distribution of this part of the 
\object{Magellanic Stream}. 
The cloud located at ($l$,$b$) = (73.5\degr, --68\degr) has a horse-shoe like shape and asymmetric 
line-profiles with wings in direction towards less negative velocities. 
The Stream fans out into numerous curved filaments and isolated clouds north of this cloud. 
Sembach et al. (\cite{sembach2}) detected \ion{O}{vi} absorption associated with the \object{Magellanic Stream}. 
They detected also \ion{O}{vi} absorption at similar velocities along sight-lines where no \ion{H}{i} 
emission has been detected to far. 
These detections indicate that the ionized component of the \object{Magellanic Stream} is more extended than 
the neutral gas. The total \ion{H}{i}~mass of this part of the Stream, $b \ge$ --70\degr, 
is $M$(\ion{H}{i}) = 0.9$\cdot$10$^7$~M$_\odot \left[d/55~{\rm kpc}\right]^2$. The mass of the ionized
component is not well constrained by observations.

The \ion{H}{i} gas in the very-high-velocity part of the \object{Magellanic Stream} has typical line-widths 
in the range $\Delta v_{\rm FWHM}$ = 20 -- 35~km\,s$^{-1}$ (see Fig.~\ref{streamnord}, spectra A and C). 
The cloud located at ($l$,$b$) = (94.5\degr, --54.5\degr) is of particular interest:
it shows more negative velocities than the neighboring clouds, it has a peak column density of 
$N$(\ion{H}{i}) = 6$\cdot$10$^{19}$~cm$^{-2}$ and specifically, the profile shows two distinct 
components. One with a line-width that is normal for the \object{Magellanic Stream} and one
component with a surprisingly narrow line-width. The lowest line width of 
$\Delta v_{\rm FWHM} \approx$ 4~km\,s$^{-1}$ (Fig. \ref{streamnord}, spectrum B) is observed at the 
northern edge of this cloud at ($l$,$b$) = (93\fdg6, --53\fdg7). The line-width allows to derive an 
upper limit for the temperature of $T_D$ = 21.8$\cdot\Delta v^2_{\rm FWHM}$ = 350~K. The kinetic 
temperature of the \ion{H}{i} gas must be lower, as turbulent motions also broaden the line. This 
cloud proves the existence of a cool gas phase in the \object{Magellanic Stream}.

\begin{figure*}[t!]
\includegraphics[width=17.8cm]{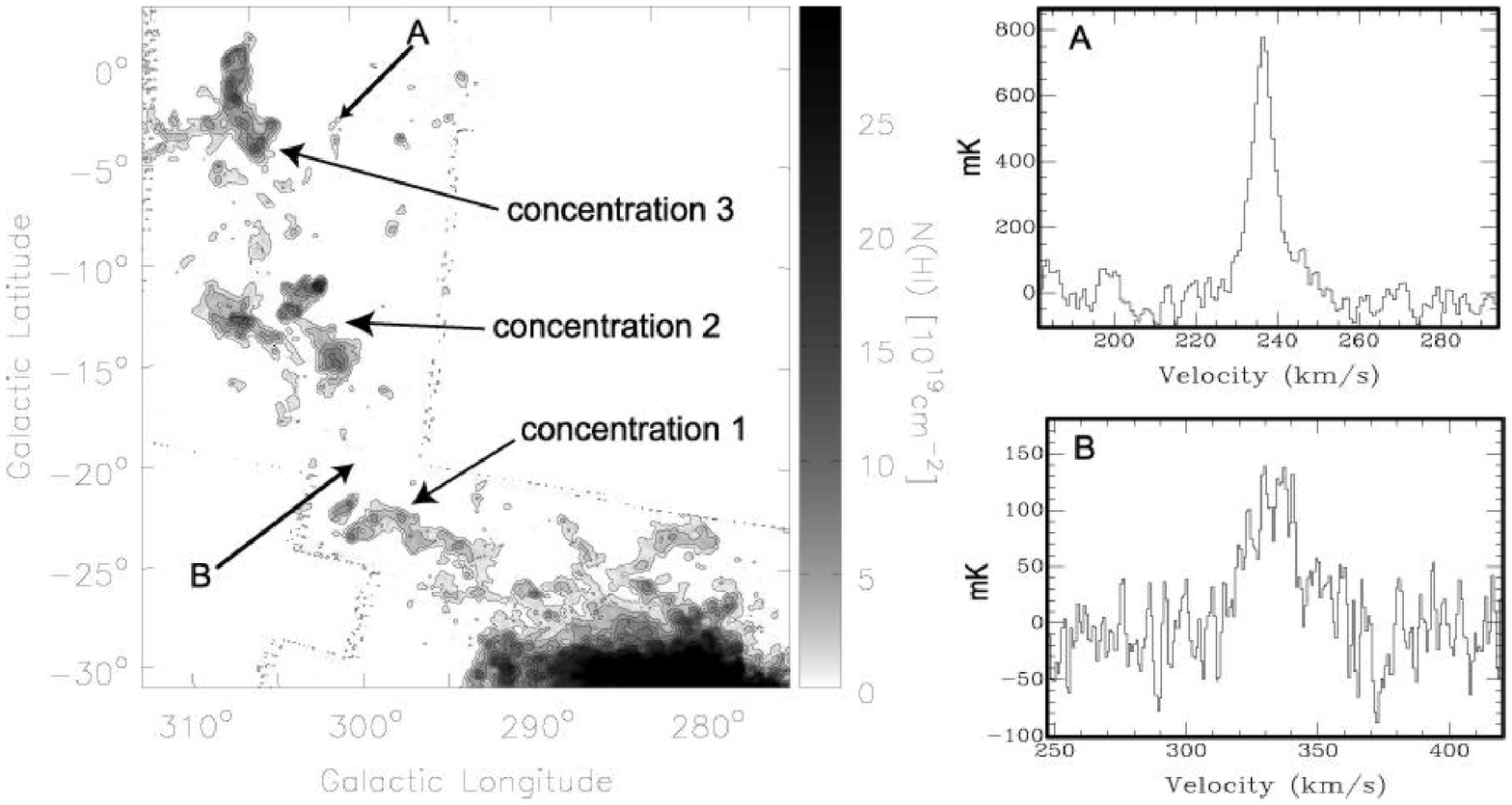}
\caption{\ion{H}{i} column density map of the very-high-velocity clouds close to the \object{LMC}
 integrated over the velocity interval 150~km\,s$^{-1}~\le~v_{\rm LSR}~\le$~420~km\,s$^{-1}$ for $b \le$ --10\degr\
 and 180~km\,s$^{-1}~\le~v_{\rm LSR}~\le$ 420~km\,s$^{-1}$ for $b >$ --10\degr\ to exclude emission from the
 Galactic Plane. 
 The contour-lines indicate \ion{H}{i} column densities of $N$(\ion{H}{i}) = 0.5$\cdot$10$^{19}$cm$^{-2}$ and 
$N$(\ion{H}{i}) = 2$\cdot$10$^{19}$cm$^{-2}$, further increasing in steps of $N$(\ion{H}{i}) = 2$\cdot$10$^{19}$cm$^{-2}$.
The spectrum A shows a compact cloud at ($l$,$b$) = (300\fdg6,--3\fdg1) with a FWHM line width of $\Delta v$ = 6~km\,s$^{-1}$.
The spectrum B, located at ($l$,$b$) = (300\fdg3,--20\fdg3), indicates that there is faint emission connecting the main concentrations of the
complex \object{LA\,I}. This spectrum is located on a line-of-sight where two observed fields overlap leading to
a $\sqrt{2}$ lower noise of this spectrum.}
 \label{LA1}
\end{figure*}

\subsection{Isolated Clouds towards the Sculptor Group}

Figure~\ref{streamseite} shows the \object{Magellanic Stream} clouds and the galaxies 
\object{NGC\,55}, \object{NGC\,300}, and \object{NGC\,7793}, having distances of 
1.6~Mpc, 2.1~Mpc, and 3.4~Mpc, respectively (Cote et al. \cite{cote}). 
These galaxies form the near part of the Sculptor group. The two galaxies 
\object{NGC\,55} and \object{NGC\,300} fall exactly onto the line-of-sight to the \object{Magellanic Stream}. 
They have radial velocities of the order of $v_{\rm LSR}$ = 100~km\,s$^{-1}$, while the 
main filament of the  \object{Magellanic Stream} has velocities in the interval
--30~km\,s$^{-1} \le v_{\rm LSR} \le$ 30~km\,s$^{-1}$\ at that position. 
The galaxies \object{NGC\,55} and \object{NGC\,300} are easy to identify, as they 
have high intensities and a large extent in velocity.
There are a number of small clouds in this region having high positive 
velocities deviating up to $\Delta v \approx$~200~km\,s$^{-1}$\ from the main filament 
of the \object{Magellanic Stream}. The enormous scatter of radial velocities near 
Magellanic longitude $L_{\rm M}$ = --50\degr\, is nicely visible in Fig.~\ref{streamseite}. 
\object{NGC\,55} is located at Magellanic longitude 
$L_{\rm M}$ = --49.2\degr\, and \object{NGC\,300} at $L_{\rm M}$ = --47.3\degr.
The small \ion{H}{i}~clouds in this region are located in the same region on the sky 
as the galaxies \object{NGC\,55} and \object{NGC\,300} showing similar velocities. 
Mathewson et al. (\cite{mathewson2}) suggested that these clouds are located within the 
Sculptor group, but Haynes \& Roberts (\cite{haynes2}) found that the overall distribution and 
kinematics of these clouds do not fit to the Sculptor group. Numerous possible origins of these 
clouds have been discussed in the literature (see Putman et al. \cite{putman2} and references therein).
Recently, Bouchard et al. (\cite{bouchard}) argued that some of these clouds might be associated 
with the \object{Sculptor dSph galaxy} that is located close to the \object{Magellanic Stream} at 
($l$,$b$) = (287\fdg53, --83\fdg16).

\section{\ion{H}{i} gas in the Region of the Leading Arm}\label{leadarm}

Figure~\ref{allmag} shows, next to the Magellanic Clouds and the \object{Magellanic Stream}, 
the \ion{H}{i} column density distribution of very-high velocity clouds north of the \object{LMC}.
The very-high-velocity gas in Fig.~\ref{allmag} represents the majority of the population EP
high-velocity clouds (see Wakker \& van Woerden \cite{wvw}).

The very-high-velocity clouds can be grouped into three complexes. First, a complex 
that is located between the Magellanic Clouds and the Galactic Plane. This complex 
was discovered by Mathewson et al. (\cite{mathewson}).
Second, a $\approx$25\degr\ long filament (285\degr $\le l \le$ 295\degr, 0\degr\ $< b <$ 30\degr), 
that is part of HVC-complex WD (Wannier et al. \cite{wannier2}). 
Third, a complex (265\degr\ $\le l \le$ 280\degr, 0\degr\ $< b <$ 30\degr), that is also  
part of HVC-complex WD. 
The three complexes will be called \object{LA\,I}, \object{LA\,II}, and \object{LA\,III} henceforward.

\subsection{Complex LA\,I}\label{sectla1}

The complex \object{LA\,I} starts close to the \object{LMC} and ends at the Galactic Plane 
near $l$ = 305\fdg5. It was discovered by Mathewson et al. (\cite{mathewson}) and studied in 
more detail by Mathewson et al. (\cite{mathewson3}), Morras (\cite{morrasa}), and 
Bajaja et al. (\cite{bajaja}).

Figure~\ref{LA1} shows a \ion{H}{i}~column density map of the complex \object{LA\,I}.
The complex does not appear as one coherent stream, but comprises three main concentrations with a 
high column density and a number of faint, compact clouds. 

The first concentration forms a linear structure with several small clouds between 
($l$,$b$) = (293$^\circ$, --25$^\circ$) and (301$^\circ$,~--22$^\circ$) that is 
directly connected with the \ion{H}{i} clouds close to the \object{LMC}, both in position 
{\em and} in radial velocity (see also Fig.~\ref{mbvel}).
 The peak column density is $N$(\ion{H}{i})~=~6.9$\cdot$10$^{19}$cm$^{-2}$ and the 
 total \ion{H}{i}~mass is $M$(\ion{H}{i})~=~1.2$\cdot$10$^6$~M$_\odot \left[d/55~{\rm kpc}\right]^2$.
 The radial velocities are in the range 230~km\,s$^{-1}$~$\le~v_{\rm LSR} \le$~330~km\,s$^{-1}$. 
Typical line-widths are in the range 20~km\,s$^{-1}$~$\le~\Delta v_{\rm FWHM}~\le$~30~km\,s$^{-1}$.

The second concentration consists of three big clouds surrounded by a number of compact clouds.
This concentration is connected with the one close to the \object{LMC} by a very faint filament 
(see Fig.~\ref{LA1}). 
Two of the big clouds show a rather diffuse column density distribution with 
radial velocities of about $v_{\rm LSR} \approx$~300~km\,s$^{-1}$. 
The third big cloud has a radial velocity of $v_{\rm LSR} \approx$~245~km\,s$^{-1}$. This
cloud shows the highest column density ($N$(\ion{H}{i})~=~1.6$\cdot$10$^{20}$cm$^{-2}$) 
of \object{LA\,I}. The cloud is much more concentrated with a steep column density gradient at the border.
The total \ion{H}{i}~mass of these three clouds is $M$(\ion{H}{i})~=~3.8$\cdot$10$^6$~M$_\odot \left[d/55~{\rm kpc}\right]^2$.

The third concentration is connected with the second one by a number of faint clouds (see Fig.~\ref{LA1}). 
This concentration is a 5$^\circ$~long and 1$^\circ$~wide complex that is orientated perpendicular to the 
Galactic Plane. The radial velocities of this filament vary from 
$v_{\rm LSR} \approx$ 250~km\,s$^{-1}$ to $v_{\rm LSR} \approx$ 220~km\,s$^{-1}$\ near the Galactic 
Plane. There is a 5$^\circ$\ long low column density feature at $b \approx$~--3$^\circ$, that is 
orientated parallel to the Galactic Plane. The third part of \object{LA\,I} has a peak column 
density of $N$(\ion{H}{i})~=~1.4$\cdot$10$^{20}$cm$^{-2}$ and a total \ion{H}{i}~mass of 
$M$(\ion{H}{i})~=~4.4$\cdot$10$^6$~M$_\odot \left[d/55~{\rm kpc}\right]^2$. 
The high column density regions show line widths of about $\Delta v_{\rm FWHM} \approx$~30~km\,s$^{-1}$,
while the gas at the northern end of this section (that overlaps with the Galactic Plane) has 
relatively low line-widths  in the range 5~km\,s$^{-1}$~$\le~\Delta v_{\rm FWHM}~\le$~10~km\,s$^{-1}$. 
A line-width of $\Delta v_{\rm FWHM} \approx$ 5~km\,s$^{-1}$ gives an upper limit to 
the kinetic temperature of $T_{\rm D}$~=~545~K. The clouds near the Galactic Plane 
appear to be colder or less turbulent than other clouds in \object{LA\,I}. 

The steep column density drop off and the narrow line widths close to the Galactic Plane can be 
interpreted as indication for a ram-pressure interaction with an ambient medium 
(Quilis \& Moore \cite{quilis}; Konz et al. \cite{konz}). The radial velocity of the fastest 
(and therefore most distant) component of the Galactic Disk at this position ($l \approx$ 305\degr) 
is $v_{\rm LSR}$ = 122~km~s$^{-1}$. Extrapolating the rotation curve from Brand \& Blitz (\cite{brand}) 
yields a distance of $d \approx$~30~kpc to the outermost spiral arm of the Milky Way (McClure-Griffiths 
et al. \cite{mcclure}). It is expected that some diffuse, low density gas exists even beyond 
$d \approx$~30~kpc.
Recent n-body simulations predict a distance between 30 to 40~kpc to the \object{Leading Arm} 
in this region (see Fig.~2 of Gardiner \cite{gardiner} or Fig.~11 of Yoshizawa \& Noguchi \cite{yoshizawa}). 
The comparable distances suggest a ram-pressure interaction between the \object{Leading Arm}
and the gas in the very outer regions of the Galactic Disk.

The total \ion{H}{i}~mass of the three main concentrations of \object{LA\,I} is 
$M$(\ion{H}{i})~=~9.4$\cdot$10$^6$~M$_\odot \left[d/55~{\rm kpc}\right]^2$.
The main concentrations of \object{LA\,I} are accompanied by a number of small clouds with similar 
velocities (see Fig.~\ref{LA1}). 

There are some compact and isolated clouds located between 294$^\circ$ $\le l \le$ 298$^\circ$ 
that have radial velocities in the interval 300~km\,s$^{-1}$\ $\le v_{\rm LSR} \le$ 350~km\,s$^{-1}$.
The region between the Galactic Plane and $b$ = --20\degr\ and $l \le$ 295\degr\  
was not covered by this survey (see Fig.~\ref{allmag}). Putman et al. (\cite{putman4}) found a number 
of compact clouds at similar velocities in this region using HIPASS data.
The radial velocities of these clouds are comparable to those observed in the \object{LMC} and the 
complex \object{LA\,I}I, but they do not fit to the neighboring gas of \object{LA\,I}.

\subsection{Complex LA\,II}

\begin{figure*}[p]
\includegraphics[width=17.0cm]{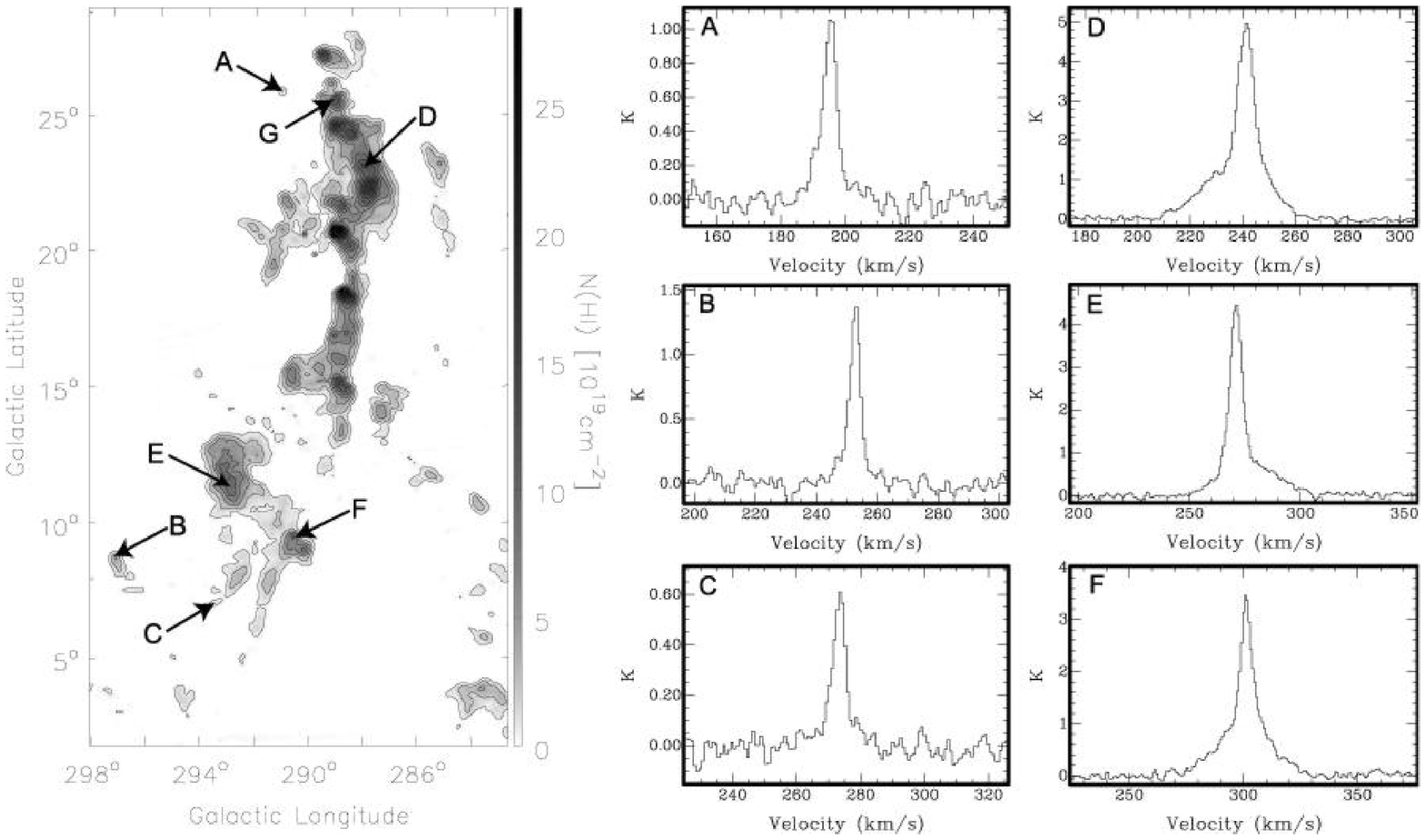}
 \caption{\ion{H}{i} column density map of the very-high-velocity clouds  in the complex \object{LA\,II}
 integrated over the velocity interval 175~km\,s$^{-1}~\le~v_{\rm LSR}~\le$~325~km\,s$^{-1}$.
 The contour-lines indicate \ion{H}{i} column densities of $N$(\ion{H}{i}) = 0.5$\cdot$10$^{19}$cm$^{-2}$ and 
$N$(\ion{H}{i}) = 2$\cdot$10$^{19}$cm$^{-2}$, increasing in steps of $N$(\ion{H}{i}) = 2$\cdot$10$^{19}$cm$^{-2}$.
The positions of the exemplary spectra A to F are ($l$,$b$) = (290\fdg9,26\fdg1), (297\fdg3,8\fdg9), (293\fdg6,7\fdg1), 
(287\fdg7,23\fdg3), (293\fdg1,12\fdg1), and (290\fdg6,9\fdg7), respectively. The spectra A, B, and C show
narrow lines with a FWHM line width of $\Delta v$ = 5.3~km\,s$^{-1}$, $\Delta v$ = 4.1~km\,s$^{-1}$, and 
$\Delta v$ = 4.8~km\,s$^{-1}$, respectively. The spectra D, E, and F show examples of spectra with two
components: a narrow and a broad line. The Gaussian decomposition yields line widths for the two
components of $\Delta v$ = 6.7~km\,s$^{-1}$ and $\Delta v$ = 28.4~km\,s$^{-1}$ for spectrum D, 
$\Delta v$ = 6.4~km\,s$^{-1}$ and $\Delta v$ = 29.0~km\,s$^{-1}$ for spectrum E, and
$\Delta v$ = 5.4~km\,s$^{-1}$ and $\Delta v$ = 26.2~km\,s$^{-1}$ for spectrum F. 
In some clouds of this complex the two components show slightly different velocities (e.g. spectra D and E).}
\label{LA2}
\vspace{0.3cm}
\includegraphics[width=17.0cm]{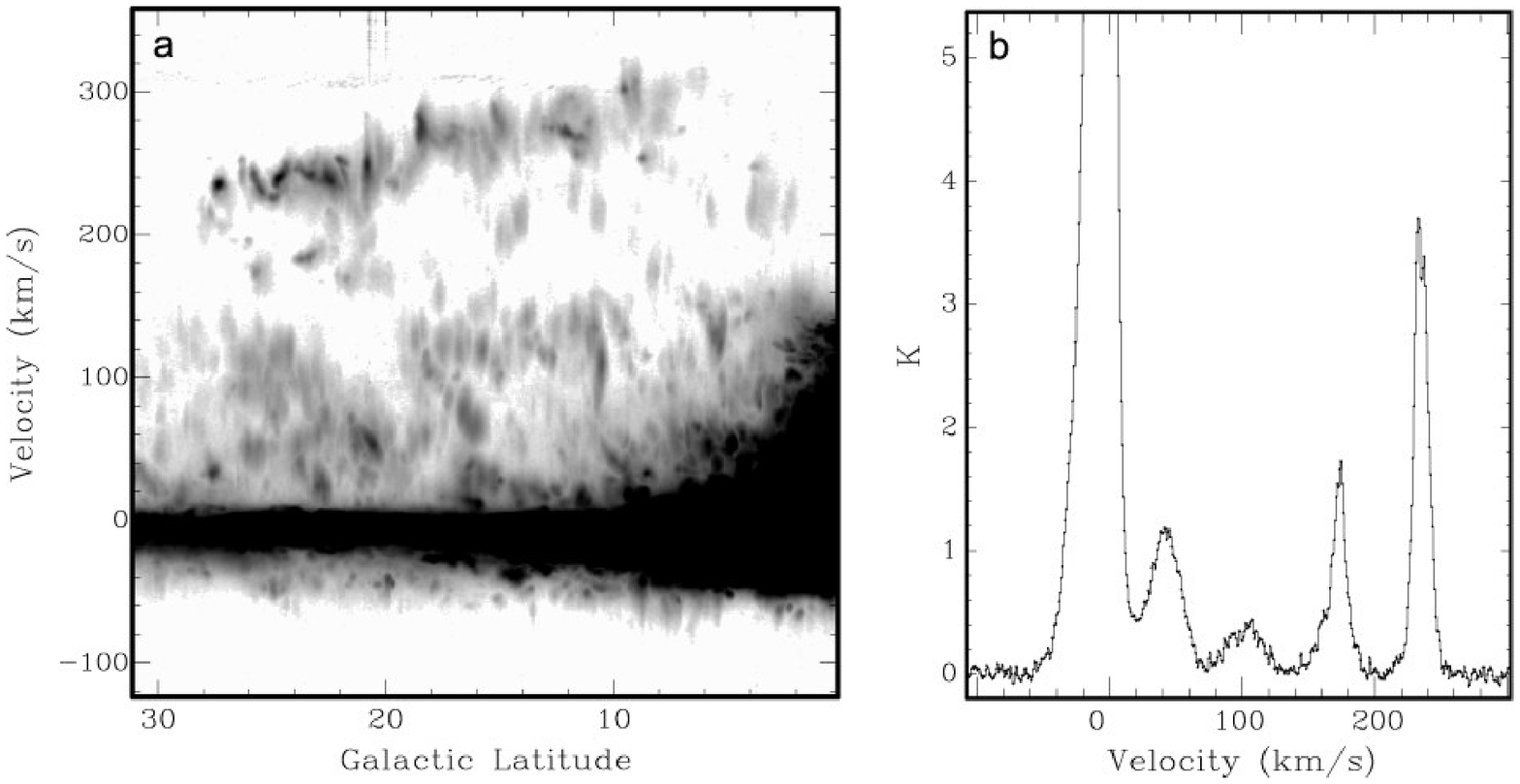}
\caption{{\bf a)} Peak intensity position/velocity map of the complex \object{LA\,II}, representing the 
maximum brightness temperature along the Galactic Longitude axis. The grey-scale has a square root scaling from 
$T_{\rm B}$ = 0~K (white) to 10~K (black). The complex \object{LA\,II} appears as a continuous, linear feature 
with a velocity gradient from about $v_{\rm LSR}$ = 310~km\,s$^{-1}$ close to the Galactic Plane to
$v_{\rm LSR}$ = 230~km\,s$^{-1}$ at its northern end. {\bf b)} An example spectrum towards ($l$,$b$) = (288\fdg78,26\fdg65)
that is marked with a ``G'' in Fig.~\ref{LA2}. The spectrum shows multiple components covering the entire 
velocity interval between $v_{\rm LSR}$ = --50~km\,s$^{-1}$ and $v_{\rm LSR}$ = +260~km\,s$^{-1}$.}
 \label{laiischnitt}
\end{figure*}

Complex \object{LA\,II} is located at 285\degr $\le l \le$ 295\degr, 0\degr\ $< b <$ 30\degr\ (see Fig. \ref{LA2}).
It is part of HVC-complex WD (Wannier et al. \cite{wannier2}) and has been studied by several groups, 
e.g. Morras \& Bajaja (\cite{morrasbajaja}), and Wakker et al. (\cite{wakker2}).

The complex \object{LA\,II} consists of two parts. The first part is a 15\degr\ long, linear 
filament between 13\degr\ $\le b \le$ 28\degr. It comprises 11 high column density clouds 
($N$(\ion{H}{i})~$\ge$~1$\cdot$10$^{20}$cm$^{-2}$) that are not resolved by the Parkes beam. 
The highest column density of $N$(\ion{H}{i})~=~2.9$\cdot$10$^{20}$cm$^{-2}$, is located at 
($l$,$b$)~=~(288.8\degr,~20.8\degr).
The second part consists of a complex of clouds in the area 290\degr\ $\le l\,\le$~295\degr\ and 
5\degr~$\le~b~\le$~13\degr\ (Fig. \ref{LA2}). Both parts are connected by a bridge of very faint gas. 
The total \ion{H}{i} mass of \object{LA\,II} is $M$(\ion{H}{i})~=~1.1$\cdot$10$^7$~M$_\odot \left[d/55~{\rm kpc}\right]^2$.

The radial velocities of complex \object{LA\,II} start at $v_{\rm LSR}~\approx$ 310~km\,s$^{-1}$ close to the 
Galactic Plane and decrease to $v_{\rm LSR}~\approx$ 230~km\,s$^{-1}$ towards $b \approx$ 30\degr\ 
(Fig. \ref{laiischnitt}a). This velocity gradient is slightly lower in the Galactic-standard-of-rest frame: 
from $v_{\rm GSR}~\approx$ 105~km\,s$^{-1}$ to $v_{\rm GSR}~\approx$ 45~km\,s$^{-1}$.

Close to the main filament there are a number of compact clouds with line widths down to 
$\Delta v_{\rm FWHM} \approx$ 4~km\,s$^{-1}$ (Fig.~\ref{LA2}, spectra A, B, and C), indicating gas with 
temperatures below 350~K. Some clouds show two components in the spectra, a narrow and a broad line, 
indicating a warm and a cold gas-phase (Fig.~\ref{LA2}, spectra D, E, and F). The broad component shows 
line widths of the order of $\Delta v_{\rm FWHM} \approx$ 25~km\,s$^{-1}$, indicating gas with temperatures 
below $T_{\rm D} \le$ 14\,000~K, while the typical line widths of the cold component are of the order of 
$\Delta v_{\rm FWHM} \approx$ 7~km\,s$^{-1}$, indicating gas with temperatures below $T_{\rm D} \le$ 1000~K. 
A similar morphology has also been observed for several HVC-complexes (see Wakker \& van Woerden \cite{wvwrev} 
and references therein).

Wakker et al. (\cite{wakker2}) observed the cloud in the line-of-sight to \object{NGC\,3783} at ($l$,$b$) = 
(287\fdg5, 22\fdg5) in \ion{H}{i} with high angular resolution using the ATCA.
They detected numerous small, dense, and cold cores ($T_{\rm k} <$ 500~K) that are embedded in a smoother 
\ion{H}{i} envelope. 

\begin{figure}[t]
\includegraphics[width=8.4cm]{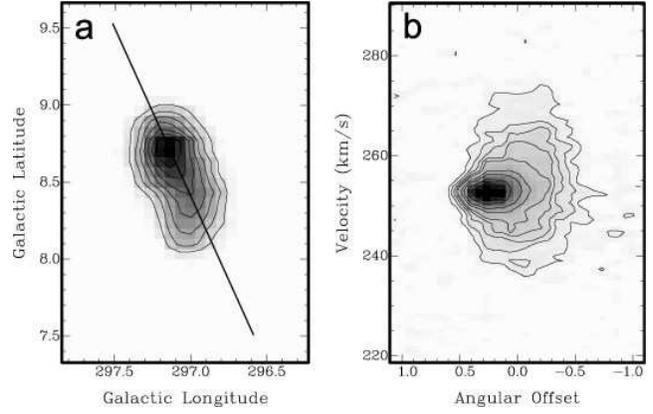}
 \caption{{\bf a)} \ion{H}{i} column density map of a head-tail cloud (cloud B in Fig.~\ref{LA2}). 
 The contour lines start at
 $N$(\ion{H}{i}) = 0.5$\cdot$10$^{19}$cm$^{-2}$ and increase in steps of $N$(\ion{H}{i}) = 0.5$\cdot$10$^{19}$cm$^{-2}$.
 The line shows the position of the slice shown in Fig. \ref{headtail}b. 
 {\bf b)} A position/velocity slice through the data-cube presented in Fig. \ref{headtail}a. 
 Positive angular offsets are in direction of increasing Galactic Latitude. The contour lines
 represent brightness temperatures of $T_{\rm B}$ = 0.1, 0.2, 0.3, 0.4, 0.5, 0.75, 1.0, 1.5, and 2.0~K.
 Near offset angle +0\fdg5 only a narrow component is visible in the spectra, near offset angle +0\fdg2
 a narrow and a broad line are detected, and near offset angle --0\fdg25 only a broad line is visible.}
 \label{headtail}
\end{figure}

Some clouds close to the Galactic Plane show a head-tail structure. These clouds have a high column density 
head with a narrow and a broad line component and a diffuse tail showing solely a broad component. 
Figure \ref{headtail} shows one example of a cloud with an asymmetric column density distribution, a 
so-called head-tail structure (Br\"uns et al. \cite{bruens1}; Br\"uns et al. \cite{bruens2}). 
The position/velocity slice through this cloud (Fig. \ref{headtail}b) shows that there is only a narrow 
component visible at the northern edge of the cloud ($\Delta v \approx$ 4~km\,s$^{-1}$), while there are 
two components detected in the middle part of that cloud and only a broad component towards the southern end
($\Delta v \approx$ 20~km\,s$^{-1}$). This morphology is suggestive of a ram-pressure interaction, where the 
diffuse gas (represented by the broad line) is stripped from the cold 
core (represented by the narrow line). This interpretation is supported by the numerical simulations from 
Quilis \& Moore (\cite{quilis}) and Konz et al. (\cite{konz}).
In the context of these simulations, this cloud would move northwards through an ambient medium. 

Figure \ref{laiischnitt}a shows that there are a number of compact clouds with radial velocities that are about
60~km\,s$^{-1}$ lower than those of the complex \object{LA\,II}. One example is a compact HVC with a diameter 
of about 40\arcmin\ centered on ($l$,$b$) = (288\fdg78, 26\fdg65). Figure~\ref{laiischnitt}b shows a spectrum 
of this line-of-sight. The compact cloud has a velocity of $v_{\rm LSR}$ = 170~km\,s$^{-1}$, approximately 
60~km\,s$^{-1}$ lower than the velocity of the \object{LA\,II}, but about 60~km\,s$^{-1}$ higher than the 
extended emission at $v_{\rm LSR}$ = 110~km\,s$^{-1}$ in this region.

The \ion{H}{i} emission around $b \approx$ 25\degr\ at $v_{\rm LSR}$ = 110~km\,s$^{-1}$ and the emission
between 10\degr\ $\le b \le$ 20\degr\ at $v_{\rm LSR}$ = 120~km\,s$^{-1}$ show a high velocity relative to 
the local gas of the Milky Way, but they are nevertheless no HVCs. Wakker (\cite{wakkerdev}) defined 
HVCs as clouds with radial velocities that deviate by at least 50~km\,s$^{-1}$ from those velocities 
expected from Galactic rotation at their position. The radial velocity of the fastest (and therefore 
most distant) component of the Galactic Disk shows similar velocities of the order 
$v_{\rm LSR} \approx$ 120~km\,s$^{-1}$.
These velocities imply a distance of $d \approx$~30~kpc (Brand \& Blitz \cite{brand}). The clouds with 
$b \ge$ 5\degr\ can therefore be interpreted as co-rotating Galactic \ion{H}{i} clouds at a distance of 
$d \approx$~30~kpc that are located at z-heights of 5 to 10~kpc above the Galactic Disk. 
The kinematical distance, however, is only an estimate as the assumption of co-rotation is not
necessarily true for these clouds. 

\begin{figure*}[t]
\includegraphics[width=17.0cm]{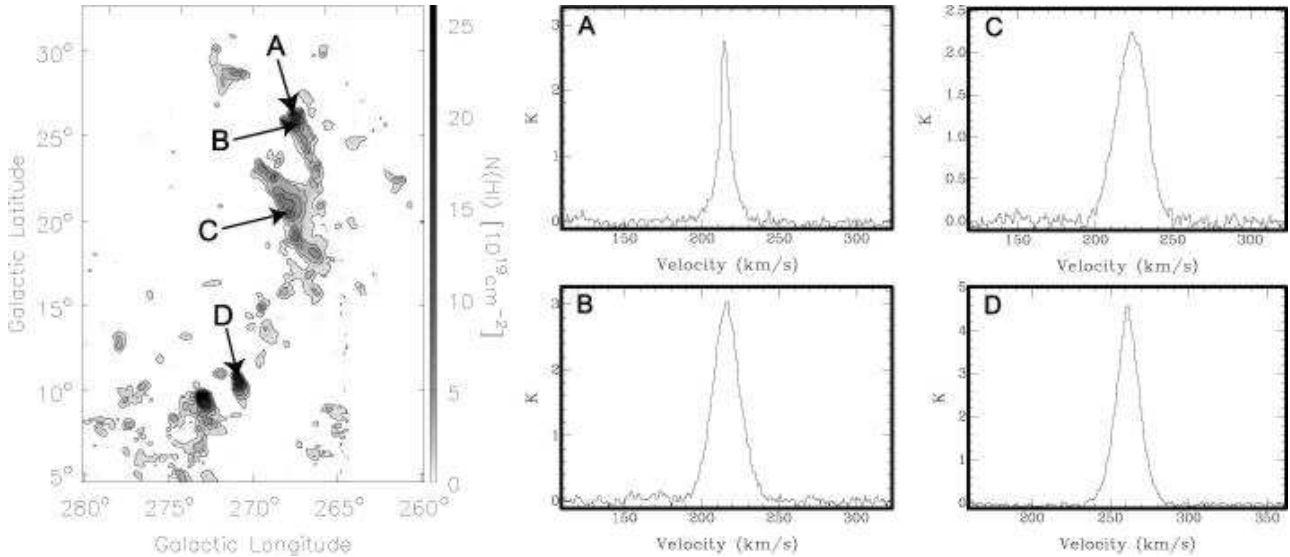}
 \caption{\ion{H}{i} column density map of the very-high-velocity clouds in the complex \object{LA\,III}
 integrated over the velocity interval 170~km\,s$^{-1}~\le~v_{\rm LSR}~\le$~300~km\,s$^{-1}$. 
 The contour-lines indicate \ion{H}{i} column densities of $N$(\ion{H}{i}) = 0.5$\cdot$10$^{19}$cm$^{-2}$ and 
$N$(\ion{H}{i}) = 2$\cdot$10$^{19}$cm$^{-2}$, increasing in steps of $N$(\ion{H}{i}) = 2$\cdot$10$^{19}$cm$^{-2}$.
The cloud at ($l$,$b$) = (267$^\circ$, 26$^\circ$) shows a steep column density gradient on one side
and an extended tail on the opposite side. Spectrum A is located at the steep gradient, ($l$,$b$) = (267\fdg5, 26\fdg7). 
It shows two components with FWHM line widths of $\Delta v$ = 4.4~km\,s$^{-1}$ and $\Delta v$ = 15.2~km\,s$^{-1}$.
Spectrum B shows the same cloud, but on the side towards the tail, ($l$,$b$) = (267\fdg2, 26\fdg0). 
The line is well represented by a single Gaussian with a line width of $\Delta v$ = 20.2~km\,s$^{-1}$. 
The spectra C and D are located at ($l$,$b$) = (267\fdg8, 20\fdg7) and ($l$,$b$) = (271\fdg0, 11\fdg1). 
They show lines with a line width of $\Delta v$ = 22.9~km\,s$^{-1}$ and $\Delta v$ = 17.3~km\,s$^{-1}$, respectively.}
 \label{LA3}
\end{figure*}

\subsection{Complex LA\,III}

Figure~\ref{LA3} shows the \ion{H}{i} column density distribution of the complex \object{LA\,III} that
is also part of the HVC-complex WD (Wannier et al. \cite{wannier2}).
The complex \object{LA\,III} is approximately parallel to \object{LA\,II}, but shifted towards lower 
Galactic longitudes (265\degr $\le l \le$ 280\degr). 
\object{LA\,III} is not as continuous as \object{LA\,II}: there are two high column density clouds close 
to the Galactic Plane ($b \approx$ 10\degr), first studied by Morras \& Bajaja (\cite{morrasbajaja}), 
and a larger cloud complex (12\degr\ by 5\degr) centered on ($l$,$b$)~=~(267\degr, +22\degr). 
These clouds have been studied in \ion{H}{i} by Giovanelli \& Haynes (\cite{giovanelli}) and 
Cavarischia \& Morras (\cite{cavarischia}). 

The two bright clouds close to the Galactic Plane at ($l$,$b$)~=~(273\fdg0, 9\fdg9) and 
($l$,$b$)~=~(270\fdg8, 10\fdg7) have peak column densities of 
$N$(\ion{H}{i})~=~2.9$\cdot$10$^{20}$cm$^{-2}$ and $N$(\ion{H}{i})~=~1.7$\cdot$10$^{20}$cm$^{-2}$, 
and mean velocities of $v_{\rm LSR}$ = 248~km\,s$^{-1}$ and $v_{\rm LSR}$ = 259~km\,s$^{-1}$, 
respectively. There are numerous low column density clouds with similar radial velocities in this region, 
especially close to the Galactic Plane. The highest radial velocity of $v_{\rm LSR}$ = 279~km\,s$^{-1}$
is observed at  ($l$,$b$)~=~(271\fdg9, 11\fdg3). The velocities are within 
210~km\,s$^{-1} \le v_{\rm LSR} \le$ 230~km\,s$^{-1}$ for the complex centered on ($l$,$b$)~=~(267\degr, +22\degr)
and $v_{\rm LSR}$ = 185~km\,s$^{-1}$ for the cloud at ($l$,$b$)~=~(271\degr, +29\degr).
The total \ion{H}{i} mass of clouds in the area of \object{LA\,III} is 
$M$(\ion{H}{i})~=~8.1$\cdot$10$^6$~M$_\odot \left[d/55~{\rm kpc}\right]^2$.

Typical line widths in this complex cover the interval 20~km\,s$^{-1}$ $\le \Delta v_{\rm FWHM} \le$ 30~km\,s$^{-1}$
(see Fig.~\ref{LA3}). These values are comparable to the broad component observed in \object{LA\,II}. 
A narrow component has been observed solely for the cloud centered at ($l$,$b$) = (267$^\circ$, 26$^\circ$).
It shows a steep column density gradient on one side and an extended tail on the opposite side, forming a 
head-tail structure. Spectrum A of Fig.~\ref{LA3} is located at the steep gradient. It shows two components 
with FWHM line widths of $\Delta v$ = 4.4~km\,s$^{-1}$ and $\Delta v$ = 15.2~km\,s$^{-1}$. Spectrum B of 
Fig.~\ref{LA3} shows the same cloud, but on the side towards the tail. The line is well represented by a 
single Gaussian with a line width of $\Delta v$ = 20.2~km\,s$^{-1}$.

The cloud at ($l$,$b$)~=~(270\fdg8, 10\fdg7) also shows an elongated structure with line widths increasing
from its northern end ($\Delta v_{\rm FWHM} \approx$ 15~km\,s$^{-1}$) to its southern end 
($\Delta v_{\rm FWHM} \approx$ 25~km\,s$^{-1}$). 

\subsection{Discussion of the \ion{H}{i} in the Leading Arm Region}\label{sectdisc}

The complexes \object{LA\,I}, \object{LA\,II}, and \object{LA\,III} have already been associated 
with the Magellanic Clouds by Mathewson et al. (\cite{mathewson}). However, two arguments have been 
used to disprove the association with the Magellanic Clouds: first, the morphological differences 
between the \object{Magellanic Stream} and the Leading Arm, especially the existence of a two-phase 
medium in the \object{Leading Arm}, and second the absence of a clear velocity gradient of the 
\object{Leading Arm} (e.g. Morras \cite{morrasa}).

The morphological differences between the neutral hydrogen in the \object{Leading Arm} and the 
\object{Magellanic Stream} include a different distribution of the gas (a coherent stream in case 
of the \object{Magellanic Stream} and clumpy filaments in case of the \object{Leading Arm}) and the 
unequal types of line profiles (the \ion{H}{i} lines show a narrow and a broad component for numerous 
clouds in the \object{Leading Arm} and only one broad component in the \object{Magellanic Stream}). 
Recent numerical simulations suggest that the distance to the \object{Magellanic Stream} is considerably 
larger than the distance to the \object{Leading Arm} (Gardiner \cite{gardiner}; Yoshizawa \& Noguchi 
\cite{yoshizawa}; Connors et al. \cite{connors}).
Moreover, the \object{Magellanic Stream} is located at high Galactic latitudes implying large distances 
from the Galactic Plane. In contrast, the \object{Leading Arm} is located at low Galactic latitudes. The 
expected distance to the \object{Leading Arm} is consistent with the distance to the outermost parts of 
the Galactic Disk (see Sect.~\ref{sectla1}). The density of the ambient medium of the \object{Leading Arm} 
is therefore expected to be much higher than the density of the medium close to the \object{Magellanic Stream}. 
Wolfire et al. (\cite{wolfire}) demonstrated that the existence of two stable gas phases depends strongly on 
the pressure of the ambient medium. The morphological differences of the gas in the \object{Leading Arm} and 
the gas in the \object{Magellanic Stream} might therefore be a product of different environmental conditions. 
Consequently, the presence of two gas phases does not argue against a  common Magellanic origin of the 
complexes \object{LA\,I}, \object{LA\,II}, and \object{LA\,III}. 

The \object{Magellanic Stream} shows a velocity gradient of $\Delta v_{\rm LSR}$ = 650~km\,s$^{-1}$ and 
$\Delta v_{\rm GSR}$ = 390~km\,s$^{-1}$ in the Galactic-standard-of-rest frame (see Sect.~\ref{streamoverall}),
while clouds in the \object{Leading Arm} do not show a clear velocity gradient (Fig.~\ref{allmagseite}). 
Recent numerical simulations of the Magellanic System (Gardiner \cite{gardiner}; 
Yoshizawa \& Noguchi \cite{yoshizawa}; Connors et al. \cite{connors}) demonstrate that a steep velocity 
gradient for the \object{Magellanic Stream} and a flat one for the \object{Leading Arm} is a natural outcome 
of simulations of the Magellanic System. The simulations do not provide a one-to-one mapping of the distribution 
of \object{Leading Arm} clouds in the position/velocity space, but the same is true for the \object{Magellanic Stream}. 
Consequently, also the second argument against a Magellanic origin of the complexes 
\object{LA\,I}, \object{LA\,II}, and \object{LA\,III} is {\em not} a conclusive presumption.

There are, however, reasons to adopt a Magellanic origin for these complexes.
The very-high-velocity complex \object{LA\,I} close to the \object{LMC} is apparently connected with the
\object{Magellanic Bridge} close to the \object{LMC}, both in position and in velocity (Fig.~\ref{LA1}). 
A common origin is therefore likely. The association of the other very-high-velocity complexes 
with the Magellanic System is less obvious. The three very-high-velocity complexes are not contiguous: 
there is a considerable gap between them. The gap in the column density distribution has a size of the 
order of 15\degr\ and the offset in radial velocity is of the order of 100~km\,s$^{-1}$. 

A number of clouds close to the Galactic Plane show a head-tail structure. These clouds have 
been detected in all three complexes \object{LA\,I}, \object{LA\,II}, and \object{LA\,III}. Recent 
numerical simulations (Quilis \& Moore \cite{quilis}; Konz et al. \cite{konz}) suggest that the 
orientation of the tail provides information on the motion relative to the ambient medium. The 
observed orientations of the tails, above and below the Galactic Plane, are consistent with clouds 
passing the Galactic Plane from southern direction as expected for clouds in a leading tidal arm of the 
Magellanic System. In contrast, clouds falling down onto the Galactic Disk should show tails pointing 
away from the Galactic Plane (e.g. Santill\'an et al. \cite{santillan}).

Further support for a Magellanic origin comes from absorption measurements. The complex \object{LA\,II} 
was observed in absorption towards the line-of-sight to the Seyfert galaxy NGC\,3783 at 
($l$,$b$) = (287\fdg5, 22\fdg5) by Lu et al. (\cite{lu}) and Sembach et al. (\cite{sembach}). 
They detected several elements and found a metallicity of 0.2 -- 0.4 solar and relative abundances 
comparable to those observed for the \object{SMC} and for the \object{Magellanic Stream} (Gibson et al. 
\cite{gibson}). Furthermore, Sembach et al. (\cite{sembach}) found molecular hydrogen, H$_2$, in absorption. 
While molecular gas has recently been detected in the \object{Magellanic Bridge} (Lehner \cite{lehner}; 
Muller et al. \cite{muller1}) and the \object{Magellanic Stream} (Richter et al. \cite{richter}), 
no molecular gas has ever been detected for ``normal'' high-velocity clouds 
(e.g. Wakker et al. \cite{wakkerco}; Akeson \& Blitz \cite{akeson}). Unfortunately, no suitable 
background sources for absorption studies are available for the other clouds in the region of the 
\object{Leading Arm}. 

The discussion in this section demonstrates that there are no reasons that contradict a Magellanic origin
of the complexes \object{LA\,I}, \object{LA\,II}, and \object{LA\,III}. On the contrary, the observed 
metallicity, the H$_2$ content, the alignment to the orbit of the Magellanic Clouds, and the direction of 
the observed head-tail structures in the \object{Leading Arm} suggest an association of the 
very-high-velocity cloud complexes \object{LA\,I}, \object{LA\,II}, and \object{LA\,III} 
with the Magellanic Clouds. Further numerical simulations including gas dynamics are
necessary to substantiate the association of these complexes with the Magellanic Clouds.


\section{Summary \& Conclusions}

The Large Magellanic Cloud, the Small Magellanic Cloud and the Milky Way 
are a spectacular set of interacting  galaxies with gaseous features that protrude 
from the Magellanic Clouds covering a significant part of the southern sky (Fig.~\ref{allmag}). 
Here, we have presented the Parkes narrow-band multi-beam \ion{H}{i}~survey of the entire Magellanic 
System. The survey comprises full angular sampling and high velocity resolution. Both are 
necessary for a detailed analysis of the dynamical evolution and the physical conditions 
of the Magellanic Clouds, and the gaseous features (\object{Magellanic Bridge}, \object{Interface Region}, 
\object{Magellanic Stream}, \object{Leading Arm}). 

A fully automated data reduction scheme was developed that generates a minimum 
median reference bandpass from the spectra of a scan. It eliminates remaining 
emission lines in the bandpass by fitting polynomials to the reference bandpass that 
are used to interpolate these regions. The routine subtracts a baseline from the 
reduced spectra and automatically detects and eliminates interference. 
The data are of high quality and allow a detailed investigation of the \ion{H}{i} 
gas in the entire Magellanic System. 

The \object{LMC} and the \object{SMC} are embedded in a common envelope called the 
\object{Magellanic Bridge}. This gas connects the two galaxies not only in position, but 
also in velocity. The observed velocity gradient of magnitude 125~km\,s$^{-1}$ between the 
\object{SMC} and the \object{LMC} can largely be explained by projection effects. 
The de-projected velocity field, using a LMC-standard-of-rest frame, 
demonstrates that the gas in the \object{Magellanic Bridge} is most likely orbiting parallel to the
Magellanic Clouds. There is a velocity gradient perpendicular to the SMC-LMC axis showing 
almost the same amplitude and orientation as the velocity field of the \object{SMC}. 
This velocity gradient and the large range of velocities  ($\Delta v \approx$ 100~km\,s$^{-1}$) 
along most lines-of-sight make a long-term stability of the gas in the \object{Magellanic Bridge} 
unlikely. A huge amount of gas emerges from all over the \object{Magellanic Bridge}, building up 
the \object{Magellanic Stream}, while the \object{Leading Arm} emerges from the \object{Magellanic Bridge} 
close to an extended arm of the \object{LMC}.

\begin{table}[t]
\caption{The peak column densities and \ion{H}{i} masses of the individual parts of the 
Magellanic System. We used a distance of 50 and 60~kpc for the \object{LMC} and the \object{SMC}, 
respectively. For the gaseous features, a distance of 55~kpc was assumed. 
The second part of the \object{Magellanic Stream} shows velocities similar to the local
gas of the Milky Way. The derived mass is therefore less well constrained.}
\begin{tabular}{lccc}
\hline
Region & $N$(\ion{H}{i})$_{\rm max}$ & $M$(\ion{H}{i}) & $v_{\rm LSR}$ range\\
 & 10$^{20}$cm$^{-2}$ & 10$^{7}$~M$_{\odot}$ & km\,s$^{-1}$\\
\hline
\object{LMC} & 54.5 & 44.1 & +165\ldots+390\\
\object{SMC} & 99.8 & 40.2 & +75\ldots+220\\
\hline
\object{Magellanic Bridge} & 16.4 & 18.4 & 110\ldots320\\
\object{Interface Region} & 5.5 & 14.9 & 60\ldots360\\
\object{Magellanic Stream}, I & 4.7 & 4.3 & 30\ldots160\\
\object{Magellanic Stream}, II & 4.8 & 4.3 & --40\ldots30\\
\object{Magellanic Stream}, III & 3.1 & 3.0 & --210\ldots--40\\
\object{Magellanic Stream}, IV & 0.6 & 0.9 & --410\ldots--210\\
\object{Leading Arm}, \object{LA\,I} & 1.6 & 1.0 & 200\ldots330\\
\object{Leading Arm}, \object{LA\,II} & 2.7 & 1.1 & 200\ldots320\\
\object{Leading Arm}, \object{LA\,III} & 2.8 & 0.9 & 160\ldots290\\
\hline
\ion{H}{i} ouside \object{LMC} and \object{SMC} & & 48.7 & --410\ldots+360\\
\hline
\end{tabular}
\label{masses}
\end{table}

The \object{Magellanic Stream} is a $\sim$100\degr\ long coherent filament passing the 
southern Galactic Pole (see Fig.~\ref{allmag}). The \object{Magellanic Stream} shows a 
considerable gradient in column density (from $N$(\ion{H}{i}) = 5$\cdot$10$^{20}$~cm$^{-2}$ to 
1$\cdot$10$^{19}$~cm$^{-2}$) and velocity ($\Delta v \approx$ 650~km\,s$^{-1}$) over its extent. 
The \object{Leading Arm} is a $\sim$70\degr\ long structure with several sub-complexes. It forms a 
counter stream to the \object{Magellanic Stream} with a relatively low velocity gradient. 
Figure~\ref{allmagseite} illustrates the distribution of the \ion{H}{i} gas in the Magellanic System 
in the position/velocity space. The data show significant differences between the clouds in the 
\object{Magellanic Stream} and those in the \object{Leading Arm}, both in the column density distribution 
and in the shapes of the line profiles. The \ion{H}{i}~gas in the \object{Magellanic Stream} is more 
smoothly distributed than the gas in the \object{Leading Arm}. Clouds in the \object{Leading Arm} quite 
often show two gas phases in form of a low and a high velocity dispersion component, while the line 
profiles in the \object{Magellanic Stream} show only a single component with a large velocity dispersion. 
The morphological differences of the \ion{H}{i} gas in the \object{Leading Arm} and in the 
\object{Magellanic Stream} could be explained by different environmental conditions. Clouds associated 
with a so-called head-tail structure, indicating an interaction with their ambient medium, were detected 
in the \object{Magellanic Stream} and the \object{Leading Arm}.

Table~\ref{masses} summarizes the observed \ion{H}{i}~masses in the Magellanic System,
assuming a distance of 55~kpc to all clouds associated with the Magellanic System and distances 
of 50~kpc and 60~kpc for the \object{LMC} and \object{SMC}, respectively.
The total \ion{H}{i} mass of the gaseous features is $M$(\ion{H}{i}) = 4.87$\cdot$10$^8$~M$_\odot \left[d/55~{\rm kpc}\right]^2$, 
if all clouds would have the same distance of 55~kpc. Approximately two thirds of these \ion{H}{i} 
clouds are located close to the Magellanic Clouds (\object{Magellanic Bridge} and \object{Interface Region}), 
and 25\% of the \ion{H}{i} gas is enclosed in the \object{Magellanic Stream}. The \object{Leading Arm}  
has a four times lower \ion{H}{i}~mass than the \object{Magellanic Stream}, corresponding 
to 6\% of the total \ion{H}{i}~mass of the gaseous arms, assuming that all 
complexes are located at a distance of 55~kpc. Putman et al. (\cite{putman2}) used HIPASS 
data to derive \ion{H}{i} masses of the Magellanic Clouds and their gaseous arms. 
They derived similar values for the \object{Magellanic Stream} and the \object{Leading Arm}, but 
significantly lower values for the Magellanic Clouds and the \object{Magellanic Bridge}. 
The lower values from the HIPASS data can easily be explained by different borders between 
the individual complexes and the observing mode, that filters out extended emission, if it 
covers the entire length of a scan.

The Parkes narrow-band multi-beam survey of the Magellanic System demonstrated 
that the amount of neutral hydrogen in the gaseous features ($M$(\ion{H}{i}) = 4.87$\cdot$10$^8$~M$_\odot \left[d/55~{\rm kpc}\right]^2$) 
is comparable to the \ion{H}{i}\ mass of the \object{LMC} ($M$(\ion{H}{i})~=~4.4$\cdot$10$^8$~M$_{\odot}$) 
and higher than the current \ion{H}{i}\ mass of the \object{SMC} ($M$(\ion{H}{i})~=~4.0$\cdot$10$^8$ M$_\odot$).
This result clarifies the strength of the interaction between the Magellanic Clouds and the 
Milky Way. The \object{SMC} would have lost more than 50\% of its original gas content during the past 
interactions, if the majority of the \ion{H}{i}\ in the gaseous features was stripped from the 
\object{SMC}, as indicated by numerical simulations (e.g. Gardiner \cite{gardiner}; 
Yoshizawa \& Noguchi \cite{yoshizawa}; Connors et al. \cite{connors}).

The future evolution of the gaseous complexes can be outlined by four scenarios:
(1) re-accretion by the Magellanic Clouds, (2) accretion by the Milky Way, (3) formation of an 
extended gaseous Galactic halo, and (4) formation of tidal dwarf galaxies.
\begin{enumerate}
\item A fraction of the gas will certainly be accreted by the Magellanic Clouds. Especially the gas 
in the \object{Magellanic Bridge} which has low velocities in the LMC-standard-of-rest frame making an 
accretion of some of this gas very likely. The infalling gas provides new fuel for star-formation.  
De Boer et al. (\cite{deboer}) found evidence for triggered star-formation in supergiant shells
located in the outer parts of the \object{LMC} and argued that it might be related to ram-pressure 
interaction with the ambient Galactic halo. The infalling \ion{H}{i} from the \object{Magellanic Bridge} 
could also be the trigger for the observed giant star-forming regions.
\item A fraction of the gas will most likely be accreted by the Milky Way providing new fuel for 
star-formation. The infall of low-metallicity gas on to the Milky Way might solve the so-called 
G-dwarf problem (see Geiss et al. \cite{geiss} and references therein). 
\item The \object{Magellanic Stream} and the \object{Leading Arm} are located far from the Magellanic 
Clouds in the outer Galactic halo. The observed head-tail structures, the detection of O{\small VI} 
(Sembach et al. \cite{sembach2}), and the detection of strong H{$\alpha$} emission 
(Weiner \& Williams \cite{weiner}; Putman et al. \cite{putman5}) associated with the \object{Magellanic Stream} 
indicate that the gaseous arms are embedded in a low-density medium. The \ion{H}{i} clouds will 
evaporate and disperse into the Galactic halo within a Gyr (Murali \cite{murali}). The total \ion{H}{i} 
mass of the gaseous arms, $M$(\ion{H}{i}) = 4.87$\cdot$10$^8$~M$_\odot \left[d/55~{\rm kpc}\right]^2$, 
corresponds to a mean density of $n_{\rm H} \approx$ 3$\cdot$10$^{-5}$cm$^{-3}$, if this gas would be 
smoothly distributed within a sphere of radius 55~kpc. The debris of past interactions between dwarf 
galaxies and the Milky Way could be the origin of the diffuse halo medium currently interacting with 
the gaseous arms of the Magellanic System.
\item Some sub-complexes of the gaseous arms have \ion{H}{i}\ masses in excess of 
$M$(\ion{H}{i}) $\sim$ 10$^7$~M$_\odot$. The total mass of these sub-complexes must be higher
as ionized or molecular hydrogen is not traced by the 21-cm line. The detection of
strong H{$\alpha$} emission associated with the \object{Magellanic Stream} demonstrates the presence
of ionized hydrogen. Helium adds about 40\% to the total mass and dark matter might also be associated
with this gas. After all, these sub-complexes provide enough material to build up a new low-mass 
dwarf galaxy in the halo of the Milky Way. A detailed analysis of the stability of these complexes
will be presented in a subsequent paper.
\end{enumerate}

\begin{acknowledgements}
     C. Br\"uns thanks the
      \emph{Deut\-sche For\-schungs\-ge\-mein\-schaft, DFG\/} (project
      number ME 745/19) as well as the CSIRO for support.
\end{acknowledgements}

\end{document}